\documentclass[12pt]{iopart}
\usepackage{graphicx}
\usepackage[T1]{fontenc} 
\usepackage{multirow}    
\usepackage{comment}
\usepackage{bbm}
\usepackage{caption}
\usepackage{cite}
\usepackage[usenames, dvipsnames]{color}
\usepackage{slashed}
\usepackage{xcolor}
\usepackage{changepage}
\usepackage{color}   
\usepackage{siunitx} 

\usepackage{hyperref}
\hypersetup{
    colorlinks=true, 
    linktoc=all,     
    linkcolor=blue,  
}

\expandafter\let\csname equation*\endcsname\relax
\expandafter\let\csname endequation*\endcsname\relax
\usepackage{amssymb}
\usepackage{amsmath}
\usepackage{amsfonts}

\newcommand{\bmtx}{\begin{pmatrix}}
\newcommand{\emtx}{\end{pmatrix}}

\newcommand{\ev}{\text{eV}}

\newcommand{\gev}{\text{GeV}}

\begin{document}

\title{Searching for the QCD Dark-Matter Axion}

\author[1]{Masha Baryakhtar}
\address{Physics Department, University of Washington, Seattle, WA 98195-1560, USA}
\ead{mbaryakh@uw.edu}

\author[1]{Leslie Rosenberg}
\address{Physics Department and Center for Experimental Nuclear Physics and Astrophysics, University of Washington, Seattle, WA 98195-1560, USA}
\ead{ljrberg@uw.edu}

\author[1]{Gray Rybka}
\address{Physics Department and Center for Experimental Nuclear Physics and Astrophysics, University of Washington, Seattle, WA 98195-1560, USA}
\ead{grybka@uw.edu}

\vspace{10pt}

\begin{indented}
\item[]\today
\end{indented}

\begin{abstract}
Proposed half a century ago, the quantum chromodynamics (QCD) axion explains the lack of charge and parity violation in the strong interactions and is a compelling candidate for cold dark matter.  The last decade has seen the rapid improvement in the sensitivity and range of axion experiments, as well as developments in theory regarding consequences of axion dark matter.  We review here the astrophysical searches and theoretical progress regarding the QCD axion.  We then give a historical overview of axion searches, review the current status and future prospects of dark matter axion searches, and then discuss proposed dark matter axion techniques currently in development.
\end{abstract}

\begin{figure}[ht]
\centering
\includegraphics[scale=0.7]{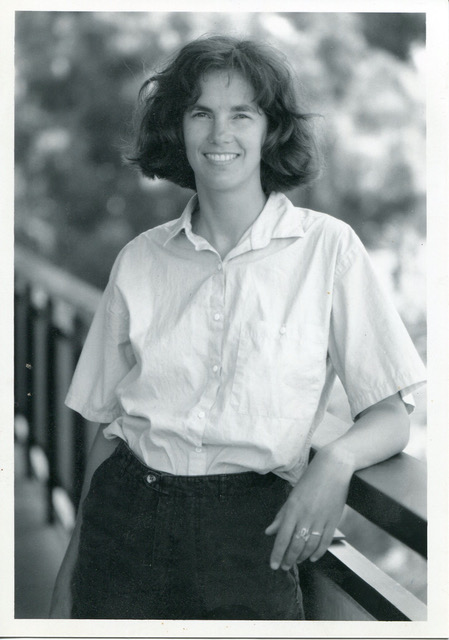}
\captionsetup{labelformat=empty} 
\caption{This review is dedicated to the memory of Ann Elizabeth Nelson.
A brilliant theoretical physicist, Ann contributed much to the theory
and phenomenology of charge and parity symmetry, including axion physics. She is loved and missed. Photo courtesy D.~B.~Kaplan.}
\addtocounter{figure}{-1} 
\end{figure}

\clearpage

\maketitle

\tableofcontents

\newpage

\section{Introduction}
\markboth{INTRODUCTION}{INTRODUCTION}
\label{sec:intro}

The particle now known as the Quantum ChromoDynamics (QCD) axion was first proposed in the late 1970s \cite{Peccei+1977,Weinberg1978, Wilczek1978}.
It originated as the low-energy manifestation of the Peccei-Quinn (PQ) mechanism  \cite{Peccei+1977,Peccei:1977ur}
devised to solve the puzzle of the undetectable smallness of the amount of charge-parity (CP) violation in the strong interactions in the Standard Model (SM) of particle physics. The CP-violating QCD field strength term has not yet been observed in exquisitely sensitive experiments and is consistent with zero. On the other hand, it is allowed by all the known symmetries of the standard model and produces physical effects such as a non-zero neutron electric dipole moment (EDM) ~\cite{tHooft:1976snw,Jackiw:1976pf,Callan:1979bg}.

This CP violation is characterized by the dimensionless parameter $\theta$,
which is the sum of the QCD topological angle 
$\mathcal{L} \supset \theta_0\, G_{\mu\nu}\tilde G^{\mu\nu} $ 
and the common quark mass phase $\theta = \theta_0 + \mathrm{arg}\,\mathrm{det M}_q$.
While the underlying physics of these two terms is unrelated, with the latter potentially obtaining contributions from CP-violating  electroweak physics,
they nonetheless appear to cancel to exquisite precision:
Among other observables,
$\theta$ contributes to the neutron EDM
$d_n \sim3\times10^{-16} \theta$~e~cm \cite{Crewther:1979pi,Pospelov:1999ha,Pospelov:1999mv}.
The latter has been experimentally bounded to be
$d_n < 1.8 \times 10^{-26}$~e~cm (90\% C.L.) \cite{Abel:2020pzs},
leading to a constraint on the CP-violating QCD field strength parameter of $\theta \lesssim 5\times 10^{-11}$. Recently, it has been noted that $\theta $ provides a contribution to electron EDM observables which currently places constraints two orders of magnitude weaker than neutron EDMs~\cite{Flambaum:2019ejc} but may provide a cross-check on the underlying physics in the event of a detection.

The QCD axion, $a$, is a pseudo-Nambu-Goldstone boson whose potential is minimized at the CP-preserving minimum~\cite{Vafa:1984xg}, up to small electroweak corrections that lie beyond the current experimental sensitivity~\cite{Pospelov:2005pr}. The QCD axion's potential~\cite{DiVecchia:1980yfw} is largely described by a single parameter,
the decay constant $f_a$, which sets its mass at zero temperature and chemical potential to~\cite{Gorghetto:2018ocs}
\begin{equation}
m_a  = 5.691(51)\,\mu\mathrm{eV}\, \left(\frac{10^{12}\mathrm{GeV}}{f_a}\right).
\label{eq:ma}
\end{equation}
The leading  self-interaction is a quartic one, with value  \cite{di_Cortona_2016}
\begin{equation}
\lambda \simeq 10^{-50}\left(\frac{10^{12}\mathrm{GeV}}{f_a}\right)^4,
\end{equation}
which, while typically negligible, can have an impact at high axion field densities such as in the early universe or near astrophysical black holes.

\begin{figure}
\includegraphics[width=\textwidth]{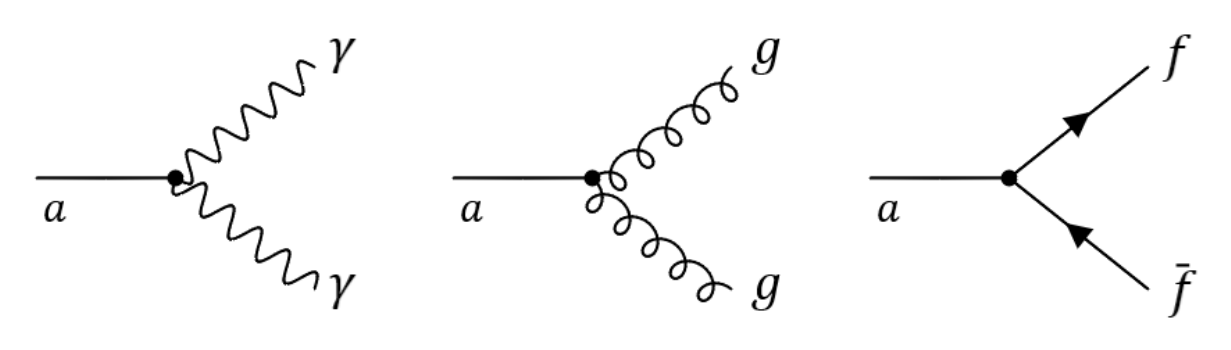}
\caption{Feynman representation of fundamental axion interactions with gluons, photons, and fermions. \label{fig:interactions}}
\end{figure}

The decay constant also  parametrizes the axion's interactions with standard-model particles at energies below the electroweak scale, including gluons, photons and fermions (Figure~\ref{fig:interactions}),
\begin{equation}
\mathcal{L} \supset  \frac{a}{f_a}\frac{\alpha_s}{8\pi} G_{\mu\nu}\tilde G^{\mu\nu} +\frac{g_{a\gamma\gamma}}{4} a F_{\mu\nu}\tilde F^{\mu\nu} + \frac{g_{a\psi}}{2}\partial_{\mu}a \bar\psi\gamma^{\mu}\gamma_5\psi,
\label{eq:lagrangian}
\end{equation}
where $G_{\mu\nu}$ ($\tilde G^{\mu\nu}$) is the gluon field-strength tensor (and its dual) and $F_{\mu\nu}$ ($\tilde F^{\mu\nu}$) is the electromagnetic field strength tensor (and its dual).
The coupling constants are parametrized as
\begin{equation}
g_{a\gamma\gamma} = C_{a\gamma\gamma}\frac{\alpha}{2\pi f_a}, \qquad
g_{a\psi} = \frac{C_{\psi}}{f_a},
\label{eq:cagg}
\end{equation}
where $\alpha$ is the electromagnetic fine structure constant and  $C_{a\gamma\gamma}, C_{\psi}$ are dimensionless coefficients, typically order-one or generated at one-loop level in generic models, as discussed further in Section~\ref{sec:theory}.

By itself, the PQ mechanism does not much constrain the axion mass [Eq.~\eqref{eq:ma}], which is to say the U(1)$_{PQ}$ sym\-metry-break\-ing scale $f_{PQ}$ (related to the decay constant $f_a$ above by the details of the PQ symmetry and its breaking) is largely undetermined.
As the axion straddles the physics of weak and strong interactions, it  seemed plausible at the time of the original proposals
to consider the electroweak  or perhaps the QCD
scale as the scale for the U(1)$_{PQ}$ symmetry breaking. 
Strong-scale axions imply axion masses in the MeV range and axion couplings to normal matter and
radiation that would have likely rendered those axions easily detectable in terrestrial experiments and in astrophysical
environments---essentially a new neutral pion-like particle would have been observed.
In the early 1980's, not long after the PQ mechanism and the axion were proposed,
the weak-scale axion was also experimentally ruled out by beam-dump, accelerator, and astrophysical searches.

As a result of these early bounds, the axion scale $f_a$ was pushed far above the electroweak scale.
These extraordinarily feeble couplings to normal matter and
radiation would make the axion perhaps the most feeb\-ly-coupled particle (excepting the graviton).
Indeed, the lifetime of such an axion into its dominant two-pho\-ton decay channel is considerably long\-er-lived than even the lifetime of the proton in a GUT scenario and many orders of magnitude
longer than the age of the universe, and can reach up to $10^{77}$ years for the lightest QCD axions.
So these are very feeb\-ly-coupled particles indeed. Such axions were, perhaps derisively, called `invisible axions' owing to
what was presumed to be the impossibility of their detection \cite{DIMOPOULOS1982185,GEORGI1982123,PhysRevLett.47.402}.

A few years after the axion was first proposed as a new particle,
it also became apparent that it would provide an inevitable contribution to the energy density of the universe~\cite{Preskill:1982cy,Abbott:1982af,Dine:1982ah}. The previously `invisible'' axion with a high $f_a$ scale turned out to have properties
that make it an attractive dark-mat\-ter candidate.
Such axions would thus not only arise in a solution to the strong-CP problem, they would also address the profound question
of the nature and composition of dark matter.
And so,  the axion became one of the rare dark matter candidates to solve an unrelated theoretical issue, the strong-CP problem.

Further, the axion has more recently been found to commonly arise in high-energy extensions to the SM with non-trivial topology in extra dimensions, such as string theory~\cite{Svrcek:2006yi,Arvanitaki:2009fg}. 
The dual theoretical and experimental motivations for the QCD axion make it one of the best-motivated candidates for a new particle beyond the SM. Therefore, the detection of the axion would yield much beyond a new particle discovery: it would advance our understanding of symmetries and provide access to the highest energy scales yet observed, as well as open the door to early universe cosmology and the local structure of dark matter in our galaxy. 

The last decade has seen tremendous progress in axion physics on both the theoretical and experimental fronts. We point the reader to a number of recent reviews on the theory~\cite{Hook:2018dlk,DiLuzio:2020wdo,Ringwald:2024uds}, experimental searches~\cite{Graham:2015ouw,Irastorza:2018dyq,Sikivie:2020zpn, Berlin:2024pzi},
and astrophysics and cosmology \cite{Marsh:2015xka,OHare:2024nmr,Caputo:2024oqc} of the QCD axion and the broader parameter space
of axion-like-particles for further detail and background material; see also the PDG review~\cite{ParticleDataGroup:2024cfk} and an online repository of recent constraints~\cite{ciaran_o_hare_2020_3932430}.
This review focuses on recent experimental achievements and proposals,
as well as new theoretical developments---with a focus on the dark matter QCD axion---while providing relevant context.
We give a summary of axion theory in Section~\ref{sec:theory},
 axion dark matter cosmology in Section~\ref{sec:cosmo},
searches in astrophysical compact objects in Section~\ref{sec:astro}, and provide a detailed review of axion laboratory experiment history, recent progress with an emphasis on cavity haloscopes, and future directions toward the dark matter QCD axion in Section~\ref{sec:experiment}. We conclude and comment on the outlook for the future in Section~\ref{sec:summary}.

\section{Theory}
\label{sec:theory}

\markboth{THEORY}{THEORY}

The details of the ax\-i\-on solution to the strong-CP problem have been established for several decades. In this section, therefore, we provide an outline of the essential theory background relevant for establishing the ax\-i\-on parameter space and subsequent discussion, in particular its mass and couplings to the SM. We also comment briefly on the more recent theoretical progress.

\subsection{The Axion Mass}
\label{sec:mass}

The ax\-i\-on mass is determined by how the QCD vacuum energy depends on the CP-vio\-la\-ting $\theta$ parameter,
\begin{equation}
m_a^2  = \frac{\chi_{\mathrm{top}}}{f_a^2} + \mathcal{O}(m_{\pi}^2/f_a^2)
\end{equation}
 where $\chi_{\mathrm{top}}$, the topological susceptibility, is the second derivative of the free energy with respect to the $\theta$ angle at $\theta=0$. This is a non-per\-tur\-ba\-tive quantity in QCD and has been computed on the lattice to be $\chi_{\mathrm{top}} = 75$~MeV with a few percent uncertainty~\cite{Gorghetto:2018ocs,Bonati:2015vqz,Borsanyi:2016ksw}\footnote{The lattice calculation is in the iso\-spin-sym\-me\-tric limit and the quoted quantity is corrected to take into account deviations from $m_u=m_d$.}. Furthermore, the fact that the physical angle $\theta$ is the sum of the QCD topological angle $\bar{\theta}$ and the common quark mass phase $\mathrm{arg} \, \mathrm{det} \, M_q$ allows the $\theta$ parameter to be fully rotated into the light quark mass matrix using quark field phase redefinitions. Then at leading order in the chiral expansion all the non-derivative dependence on the axion is contained as phases in the pion mass terms, which in turn allows the use of tools of chiral Lagrangian perturbation theory with inputs from the lattice to compute the ax\-i\-on mass very precisely at zero temperature \cite{di_Cortona_2016}. Using $N_f=2$ chiral theory at NNLO with QED corrections~\cite{Gorghetto:2018ocs} gives the mass prediction of Eq.~\eqref{eq:ma}.

At  finite temperature, the topological susceptibility is an important quantity in determining the early universe evolution of the ax\-i\-on and therefore its late-time abundance.  The  ax\-i\-on begins behaving as dark matter in the early universe when the ax\-i\-on mass is of order the Hubble parameter; the ax\-i\-on mass is itself a function of the background temperature. The high-tem\-per\-a\-ture regime $\chi_{\mathrm{top}}(T)\gg T_c$ is largely well-ap\-prox\-i\-ma\-ted by the dilute instanton approximation while at lower temperatures, the lattice has been used to determine $\chi_{\mathrm{top}}(T)$.  A shallower power law with temperature would imply a larger $f_a$ coupling constant for a fixed dark matter abundance and initial misalignment angle.   While in some work \cite{Trunin:2015yda, Bonati:2015vqz}, the power law suppression was found to be a much smaller than the dilute instanton gas prediction, $\chi_{\mathrm{top}}(T)\sim T^{-8}$,  later lattice results seem to agree with the instanton gas approximation scaling at temperatures close to the QCD phase transition in pure Yang-Mills \cite{Berkowitz:2015aua,Borsanyi:2015cka} and QCD \cite{Petreczky:2016vrs,Borsanyi:2016ksw,Burger:2018fvb,Bonati:2018blm} for three light flavors, although to date calculations of the overall magnitude of the susceptibility are not yet in agreement. The lattice calculations which capture the temperature dependence $\chi_{\mathrm{top}}(T)$ over the relevant range point to a post-inflationary misalignment axion mass of $28\pm2\,\mu$eV \cite{Borsanyi:2016ksw}. Other work estimates the present uncertainties in the topological susceptibility to translate to an uncertainty in the dark matter axion mass of a factor of $2$~\cite{Dine:2017swf}.

The understanding of the QCD axion potential at finite nuclear density (or chemical potential) is far more uncertain. At densities approaching nuclear density, the potential is expected to be suppressed~\cite{Hook:2017psm} due to a reduction in the quark condensate~\cite{Cohen:1991nk}. How this suppression---and the resulting effect on the axion mass and interactions---evolves at and beyond nuclear density remains unclear but may have  implications for axion physics in dense objects such as neutron stars~\cite{Balkin:2020dsr,Kumamoto:2024wjd,Springmann:2024ret}.

\subsection{Ax\-i\-on Interactions with Standard Model Matter and Radiation}
\label{sec:couplings}

The ax\-i\-on solution to the strong-CP problem relies only on the coupling of the ax\-i\-on to the gluon field strength. However, this coupling is unfortunately the most difficult to probe quantitatively, due to the non-perturbative nature of QCD at low energies. Our inability to access free gluons and quarks and the lack of precision predictions and measurements of their low-energy manifestations---nucleons and nuclei---make this coupling a challenge to pursue. Nevertheless, in recent years there have been several detection ideas, including Casper Electric \cite{Graham:2013gfa,Budker2013,Kimball2017} and the more recent Ax\-i\-oelec\-tric effect~\cite{Arvanitaki:2021wjk,Arvanitaki:2024dev} to search for ax\-i\-on dark matter through the fundamental QCD coupling.

It was quickly realized that in typical ultraviolet models the ax\-i\-on also inherits interactions to other SM particles~\cite{Kim:1979if,Shifman:1979if,Dine:1981rt,Zhitnitsky:1980tq,Srednicki:1985xd,Georgi:1986df}. One of the more generic, and most promising for detection, is the coupling to pairs of photons. Since part of the coupling to photons comes from low-energy QCD, it is challenging to entirely tune away the ax\-i\-on interaction with photons. There are two original benchmark classes of models yielding photon coupling predictions: KSVZ (or ``hadronic'') \cite{Kim:1979if,Shifman:1979if} and DFSZ \cite{Dine:1981rt,Zhitnitsky:1980tq}, which have different field content creating the anomaly that yields the two-pho\-ton coupling.  These envelope an  expected band of interactions~\cite{Gorghetto:2018ocs},
\begin{equation}
g_{a\gamma\gamma} = \frac{\alpha}{2\pi f_a}\left(\frac{E}{N}-1.92(4)\right),
\end{equation}
with $E$ ($N$) the electromagnetic (color) anomaly of the PQ symmetry.
Searches for the photon coupling has driven most of the experimental effort in the search for the QCD ax\-i\-on, see Section~\ref{sec:experiment}.

In addition to photons, ax\-i\-ons also have a derivative interaction with nuclei $N$, with interactions at low energies of the form,
 \begin{equation}
\mathcal{L} \supset  \frac{\partial_\mu a}{2 f_a} c_N \bar{N} \gamma^{\mu}\gamma_5 N, \quad \mathrm{with} \quad N=\{n,p\},
\end{equation}
where the couplings to protons $p$ and neutrons $n$ depend on the model and have been computed using chiral EFT methods with lattice input to be~\cite{di_Cortona_2016},
 \begin{align}
\mathrm{KSVZ}&:  c_{p} = -0.47 (3), \quad\quad\quad  c_{n} = -0.02(3).\\
\mathrm{DFSZ}&: c_{p} = -0.617 + 0.435 \sin^2 \beta \pm 0.025, \quad c_{n} = 0.254 - 0.414 \sin^2 \beta \pm 0.025.
\end{align}
In the DFSZ models there are direct couplings to the quarks in the UV theory, and $\tan\beta$ characterizes the ratio of vacuum expectation values in a two-Higgs doublet model.

The typical couplings to electrons is $\sin^2\beta/3$ in DFSZ models,  and if zero at tree level in the UV is generated at loop level at low energies to be $C_{e}\sim\mathcal{O}(\alpha^2)$~\cite{Srednicki:1985xd,Georgi:1986df}.
Finally, in a CP-violating background, ax\-i\-ons can mediate new Yu\-ka\-wa-type forces~\cite{Moody:1984ba}.  These new forces are being searched for in experiments including ARIADNE~\cite{PhysRevLett.113.161801,ARIADNE:2017tdd}.

There have been  many recent theoretical developments potentially broadening the parameter space of couplings outlined above, including enhanced couplings using the clockwork mechanism~\cite{Agrawal:2017cmd}; for a detailed recent review of models see \cite{DiLuzio:2020wdo}. However, in grand unified theories, even axion-like particles are expected to have photon couplings close to or below the QCD axion line, further motivating experimental searches that target QCD axion sensitivities~\cite{Agrawal:2022lsp,Gendler:2023kjt,Agrawal:2024ejr}.

\section{Cosmology}
\markboth{COSMOLOGY}{COSMOLOGY}
\label{sec:cosmo}

Just a few years after the axion particle was identified as a solution to the strong-CP problem, it became apparent that it could easily solve another outstanding problem in the Standard Model: the absence of a cold dark matter candidate \cite{Rubin:1970zza,WMAP:2003elm}. Several groups pointed out that the so-called invisible axion---with decay constant above that constrained by stellar evolution, $f_a \gtrsim  10^{8}$~GeV~\cite{PhysRevD.36.2201,PhysRevD.36.2211,PhysRevLett.56.26}---would have a non-tri\-vi\-al cosmological abundance~\cite{Abbott:1982af,Dine:1982ah,Preskill:1982cy}.

Cosmology with an axion field goes as follows:  Initially, the temperature of the universe is above $f_a$ and particle physics is Pec\-cei-Quinn symmetric.  As the universe cools below $f_a$, Pec\-cei-Quinn symmetry is broken, and, below the scale of the QCD crossover transition, an energetically preferred value of $\theta\equiv a/f_a$ arises.  The misalignment between the initial $\theta_i$ value and the preferred minimum of the axion potential corresponds to a large energy density being stored in the axion field.  Additionally, because $\theta$ is $2\pi$ periodic, topological defects known as axion strings thread the universe upon PQ breaking and, at the time of the QCD transition, regions of different $\theta$ are bounded by domain walls.  The strings and domain walls evolve and ultimately annihilate, emitting axion radiation.  If this process takes place before the end of inflation, and PQ symmetry is not subsequently restored, then the expected number of strings and domain walls within our visible universe is negligible.  At no time is the axion field in thermal equilibrium with the rest of the SM particles, and the axion couplings are so small that axions produced via misalignment or topological defect decay persist today as dark matter, either smoothly distributed or potentially in part in residual clumps formed in the QCD phase transition.

A detailed description of axion cosmology is available in~\cite{Marsh:2015xka,OHare:2024nmr}.  Here we summarize the aspects above, with a focus on addressing the question of what astrophysical measurements of dark matter can tell us about QCD axion parameters, and discuss current developments beyond the standard axion cosmological story.

\subsection{Axion Misalignment}
\label{sec:misalignment}

If PQ symmetry is broken before inflation ends, a range of initial conditions is possible, and the energy density in axion dark matter depends on the initial value $\theta_i$ in the patch from which our universe originated~\cite{Linde:1987bx}.  If PQ symmetry is broken after inflation ends, then all possible $\theta_i$'s are populated in our observable universe and the initial energy density arises from the average axion field misalignment oscillating around its minimum value. This gives \(\Omega_a h^2 \approx 0.12 \left(28\pm 2~\mu\mathrm{eV}/m_a \right)^{1.165}\), averaged over a flat initial $\theta$ distribution in the $(-\pi,\pi)$ range\footnote{The `effective' initial misalignment for the QCD axion potential, $\sqrt{\langle\theta_i^2\rangle}\sim 2.155$, deviates slightly from the $\sqrt{\langle\theta_i^2\rangle}=\pi/\sqrt{3}$ result for a cosine potential~\cite{di_Cortona_2016,Borsanyi:2016ksw}.}. Calculations with instanton and chiral EFT techniques suggest a larger uncertainty of $\sim25$-$45~\mu\mathrm{eV}$~\cite{Dine:2017swf}. In the post-inflationary case, axions from string and domain wall decay also add to the misalignment contribution which likely shifts the prediction to higher masses; see the following section. This scenario can be more difficult to implement in some string axion models due to the nonperturbative nature of the topological defects~\cite{March-Russell:2021zfq,Benabou:2023npn,Reece:2024wrn}, although see recent constructions~\cite{Petrossian-Byrne:2025mto,Loladze:2025uvf}. 

If PQ symmetry is never restored after inflation ends, the vacuum misalignment scenario estimates the dark matter density as \(\Omega_a h^2 \approx 0.12 \left(7~\mu\mathrm{eV}/m_a \right)^{1.165} \theta_i^2\), where \(\theta_i\) is an arbitrary initial angle randomly seeded by primordial fluctuations in the range $(-\pi,\pi)$; note this scaling is modified for $\theta\sim\pi$ due to nonlinearities in the axion potential~\cite{Arvanitaki:2019rax}.

\subsection{Axion String Networks}
\label{sec:axionstrings}

The axion arises from a spontaneously broken global symmetry, $U(1)_{\rm PQ}$. If this symmetry is broken after the end of inflation, topological defects which form during the phase transition can be relevant to axion cosmology and final dark matter abundance~\cite{Sikivie:1982qv,Vilenkin:1982ks,Vilenkin:1984ib,Davis:1986xc}.

The equations of motion for the axion and radial fields admit a string solution where the phase of the axion field winds around from $0$ to $2\pi$ and the PQ symmetry is restored at the center of the string. The size of the string core is set by the radial mode mass $m_\rho$, and away from the string the field is equal to its expectation value. The energy per unit length of the string in the core is of order $\mu\sim f_a^2$. Additional energy is stored in the gradient of the axion field away from the string core. This energy is logarithmically divergent along the direction away from the string,  in contrast to the case of a broken {local} symmetry, in which the variation of the field outside the string core is pure gauge and carries no energy. The divergence is naturally regulated by the nearest string of opposite orientation, the string loop radius, or the size of the Hubble patch. 

\begin{figure}[t]
\centering
\includegraphics[trim={0.2cm 8.7cm 1.6cm 0}, clip, width=\textwidth]{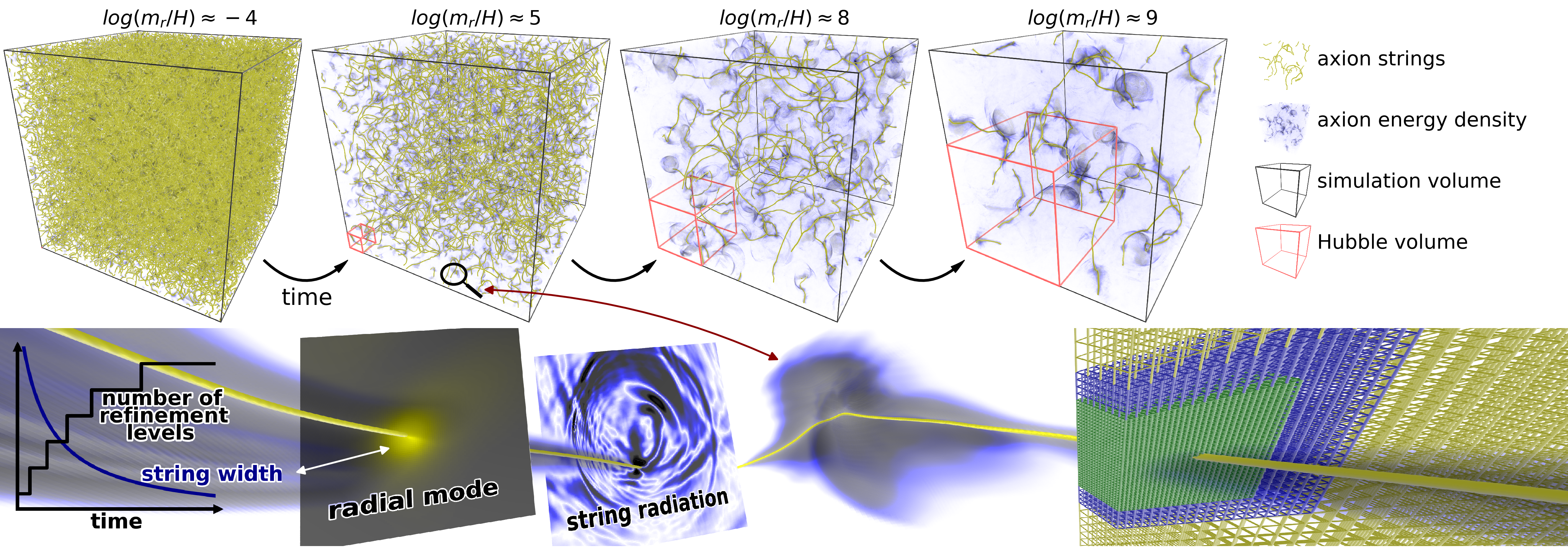}
    \caption{Illustration of the evolving axion string network in the background of an expanding universe. The growing Hubble volume is shown in red. Figure from~\cite{Buschmann:2021sdq}.}
    \label{pfig1}
\end{figure}

The dynamics of the string network in an expanding universe, illustrated in fig.~\ref{pfig1}, are complicated but can be qualitatively understood as being driven by two processes. The string length and thus energy density per Hubble volume increase as more strings enter a given Hubble volume due to the expansion of the universe. On the other hand, the strings can intersect and recombine with one another, and form loops and high-curvature kinks;  they already have high curvature components due to the random formation process which is uncorrelated across Hubble patches. High curvature strings and loops radiate axions, radial modes, and to a lesser extent gravity waves (GWs)~\cite{Vachaspati:1984gt,Figueroa:2020lvo,Chang:2021afa,Gorghetto:2021fsn}.  
Since a higher density of strings results in more crossings and radiation, the network is believed to reach a scaling solution of $\mathcal{O}(1)$ strings per Hubble volume, assuming efficient recombination and radiation of the strings, as the length of strings entering the Hubble patch is converted into axion radiation. The total energy per unit length of the string is  given by $\mu \sim \pi f_a^2 \log(\eta m_\rho/H)$, where $H^{-1}$ is the Hubble length, $m_\rho^{-1}$ the size of the string core  and $\eta$ is an $\mathcal{O}(1)$, potentially slow\-ly-time-vary\-ing number which captures the number of strings per Hubble volume~\cite{Gorghetto:2018myk}. A larger number of strings per volume results in a larger amount of axion dark matter at late times.

These radiated axions have momenta  distributed between the effective infrared cutoff, which is provided by approximately the Hubble scale $H$, and the effective ultraviolet cutoff set by the string width $m_\rho$. For momenta  between these two scales the radiation spectrum is expected to follow a power-law, 
$\propto (k/H)^{-q} \quad \mathrm{for  } \quad H  \ll k \ll m_\rho$ and a scaling index $q$. The axions  then redshift,  acquire mass during the QCD crossover transition---during which domain walls form and annihilate and potentially produce more axion radiation---and contribute to the dark matter abundance.

Most recent simulations have established the logarithmic growth of the number of strings per Hubble volume~\cite{Gorghetto:2018myk,1906.00967,Gorghetto:2020qws,Buschmann:2021sdq,Saikawa:2024bta,Kim:2024wku,Benabou:2024msj}, first observed  in~\cite{Gorghetto:2018myk}.  A  remaining source of uncertainty is the momentum spectrum of axions radiated by the strings, as this dictates the number of axions (and in the end the axion dark matter abundance). The extraction of the power law is uncertain since the UV and IR cutoffs must be far enough apart to allow for a robust extraction of the power law component of the momentum spectrum, and choices related to the cutoff and grid spacing are subject to numerical errors~\cite{Gorghetto:2020qws,Buschmann:2021sdq,Saikawa:2024bta,Benabou:2024msj}.

 It has been argued the the spectrum should be scale invariant ($q=1$) based on observations of rapidly collapsing string loops~\cite{Harari:1987ht,Hagmann:1998me}, or peaked in the IR ($q>1$), based on the similarities of the dynamics of axion strings to local strings at late times~\cite{Davis:1986xc,Battye:1993jv,Battye:1994au}. It appears unlikely to be UV dominated. It has also been observed that the power law itself may evolve logarithmically with time as the universe expands, $q \sim q_0 +q_1\log(m_\rho/H)$~\cite{Gorghetto:2018myk,Gorghetto:2020qws}. Even a small positive coefficient for the logarithmic term would lead to an IR-dominated spectrum at late times and predict a large axion
 number density and an axion mass at the meV scale from the string abundance~\cite{Gorghetto:2020qws}. Other groups have observed evolution consistent with the logarithmic growth of the power law, predicting an axion dark matter energy density dominated by string decays~\cite{Saikawa:2024bta,Kim:2024wku}; see also~\cite{Saikawa:2024bta} for a discussion of different scaling laws with time.  On the other hand, a momentum spectrum constant over the evolution, extracted from recent simulations using  adaptive mesh refinement techniques with large dynamic range, gives an  axion string contribution which only slightly dominates the misalignment value, resulting in an estimated axion mass of tens of $\mu$eV~\cite{Buschmann:2021sdq,Benabou:2024msj}.

It is evident that there is considerable uncertainty in the calculation of the energy density of axions from string and domain wall decay.  The main difficulty comes in the huge dynamical range of the problem. Placing the evolution of the complex scalar field on a grid, one must simultaneously have a small enough spacing $d$  in order to accurately resolve string core dynamics, $d\ll m_\rho^{-1}$, and large enough ``box'' volume to capture cosmic string evolution and range of momentum scales, $L \gg H^{-1}$. By the time the string network is destroyed at the QCD transition, the number of grid points to evaluate in principle should reach $(L/d)^3\gg (m_\rho/H)^3 \sim e^{200}$, an unimaginable task. Thus any simulation results have to be extrapolated from smaller grid sizes which only capture the early stages of string network evolution.

Further uncertainties arise due to the complex nature of the formation and collapse of the domain wall network. The domain walls form when the axion potential turns on during the QCD crossover transition. The value of the axion $\theta$ angle jumps from $0$ to $2\pi$ at the domain wall. The system then consists of strings connected by domain walls, as  well as closed or infinite wall surfaces without strings. The dynamics then depend on the vacuum structure, i.e., the number of minima $N$ in the axion potential. A single minimum corresponds to unstable domain walls, where as a higher number of minima can result in a frustrated network which is stable; to make $N>1$ viable, a  breaking of the PQ symmetry is required,  small enough to not reintroduce the strong-CP problem.  The contribution of this process to dark matter axion abundance is not yet well understood but may well be larger than the string production~\cite{Gorghetto:2020qws,Gorghetto:2022ikz,Benabou:2024msj}.

\subsection{Inferences on the Preferred Axion Mass from Dark Matter Density}
\label{sec:abundance}

In principle, with perfect understanding of the mechanisms above, the energy scale $f_a$ and thus the axion mass could be inferred from density of dark matter today.  Unfortunately, this has challenges in both the case where PQ symmetry is broken before and after inflation.  In the pre-in\-fla\-tion\-ary case, the dark matter density depends strongly on the initial $\theta$ in our universe.  A typical, $\mathcal{O}(1)$, misalignment angle suggests an axion dark matter mass scale $ \sim 1$--$30~\mu\mathrm{eV} $, the focus of resonant cavity experiments since the 1980s.  However, misalignment angles exponentially close to $0$ or $\pi$ can yield axion dark matter mass scales ranging from peV to meV~\cite{Linde:1987bx,Tegmark:2005dy,Arvanitaki:2019rax}. 

In the post-in\-fla\-tion\-ary PQ symmetry breaking case, there is in principle a unique dark matter abundance prediction as a function of axion mass. To find it, it is important to establish the dynamics of the axion string network; both the number of strings per Hubble volume as a function of time, and the spectrum of axions that are radiated, all of which suffer from the uncertainties mentioned above.  Over the last several decades, both numerical and analytical approaches have been applied and reached a range of conclusions, from the strings contributing a smaller or similar number of axions relative to the initial misalignment value, which would point to a scale of $f_a\sim 2\times10^{11}$~GeV \cite{Harari:1987ht,Hagmann:1990mj,1708.07521,1906.00967,Dine:2020pds,Buschmann:2021sdq,Benabou:2024msj}, to the contribution of the string network dominating the axion production by about an order of magnitude, resulting in an axion mass necessary to achieve the right relic abundance on the order of 0.5\,meV~\cite{Davis:1989nj,Battye:1993jv,Battye:1994au,Gorghetto:2018myk,Gorghetto:2020qws,Hoof:2021jft,Kim:2024wku}.

In summary, we feel it will likely never be possible to fully simulate the evolution of the complex dynamics giving rise to axion dark matter production from strings and domain walls. However, there have been new results achieved due to the combination of increased computational power, improved numerical techniques such as adaptive mesh refinement achieving impressive dynamic range~\cite{Buschmann:2021sdq, Benabou:2024msj}, and a connection with analytical arguments which can inform the extrapolation to unachievably high dynamic ranges.  Recent advancements in the scale of these simulations suggest higher axion masses from string production than those predicted by the vacuum misalignment mechanism alone. Buschmann and Benabou et.\,al. find a mass range of $40$--$180\,(45$--$65)\,\mu\mathrm{eV} $ for KSVZ~\cite{Buschmann:2021sdq,Benabou:2024msj}. Spectra consistent with IR-dominated momenta at late times yield larger masses: Gorghetto et.\,al.~\cite{Gorghetto:2020qws} find $m_a \sim 500~\mu\mathrm{eV} $ for the KSVZ model (domain wall number $N=1$) and $ \sim 3500~\mu\mathrm{eV} $ for the DFSZ model ($N=6$); Saikawa et.\,al.~\cite{Saikawa:2024bta} report a best fit mass from $95$--$450~\mu\mathrm{eV} $, and Kim et.al. $\sim 420$--$470\,\mu$eV~\cite{Kim:2024wku}. Klaer and Moore, on the other hand, find a lower value $26.2\pm 3.4\,\mu$eV~\cite{1708.07521} from a different string implementation than the other approaches.
A consensus of the community could provide interesting input to the search for axion dark matter. As there are likely to be additional contributions from domain wall collapse~\cite{Gorghetto:2020qws,Benabou:2024msj}, these mass ranges should be taken as lower bounds for post-inflationary dark matter axions.

It is important to keep in mind of course that the pre-in\-fla\-tion\-ary scenario is  perfectly consistent with everything we know about axions and the early universe and can significantly broaden the range of axion masses for which the axion is a motivated dark matter candidate.

\subsection{Additional Cosmological Signatures of Axion Dark Matter}

There are cosmological bounds on the axion parameter space, which apply to the QCD axion as well as more general axion-like particles. In the pre-in\-fla\-tion\-ary scenario, the axion is present as a light scalar field during inflation, which undergoes its own fluctuations independently from the inflaton, leading to contributions to isocurvature fluctuations~\cite{Turner:1990uz,Hertzberg:2008wr}. The amplitude of fluctuations is proportional to $(H_I/f_a)^2$ where $H_I$ is the value of the Hubble scale during inflation, and  are constrained to be small by measurements of the Cosmic Microwave Background (CMB)~\cite{Planck:2018jri}. This results in the requirement that if $f_a>H_I$ and PQ symmetry is not restored after inflation, then $f_a\gg H_I$, that is the scale $f_a$ has to be far enough above the scale of inflation to suppress isocurvature contributions. 

Regardless of the production mechanism of axion dark matter, a small abundance of relativistic axions can be produced in the early universe and contribute to the amount of radiation today, typically parametrized as an equivalent increase in the number of neutrino species, $N_{\textrm{eff}}$. At high enough temperatures, the abundance of axions produced can be significant enough to place constraints on axion parameter space from the non-ob\-ser\-va\-tion of extra relativistic radiation in the BBN, CMB, and large scale structure analyses~\cite{Chang:1993gm,Weinberg:2013kea,Brust:2013ova,Graf:2010tv,Masso:2002np,Salvio:2013iaa,Baumann:2016wac,Notari:2022ffe}. Future CMB surveys could probe $N_{\textrm{eff}}$ to a high enough precision to exclude or observe axions with a low enough $f_a$. These constraints assume a high reheating scale, but would otherwise be competitive with constraints from the cooling of astrophysical objects~\cite{Baumann:2016wac}. The couplings of axions to fermions below the electroweak symmetry breaking scale, on the other hand, are dominated by low energies, and are therefore insensitive to the scale of inflation, but at the moment do not produce competitive bounds~\cite{Green:2021hjh,Caloni:2022uya}.

\subsection{Modifications of Axion Cosmology}

There have been several recent developments in the theoretical aspects of the cosmological history of axions that can affect the prediction of axion DM abundance today as a function of its mass and decay constant. These fall into two categories: increased abundance due to early universe dynamics, or decreased abundance through  depletion mechanisms. In addition, as in any dark matter model, the cosmological history will affect the final axion abundance; for example an era of early dark matter domination can enhance axion abundance as well as axion substructure \cite{Blinov:2019jqc, Blinov:2019rhb }.

One way to achieve an axion abundance larger than the minimal prediction is to place the axion field close to the top of its potential, delaying the start of oscillations and red-shift\-ing.
In the large misalignment mechanism~\cite{Arvanitaki:2019rax}, it was pointed out that large initial field values (exponentially close to the maximum of a cosine potential, or generically for potentials which are relatively `flatter' at large field values) can result in a parametric increase in the dark matter density. In addition, the larger self-interactions which result from probing the non-quadratic part of the potential give rise to interesting dynamics, including parametric resonance of certain axion momentum modes and enhancement of structure on small scales. Several models which implement the above initial conditions have been proposed \cite{Co:2018mho,Takahashi:2019pqf,Huang:2020etx}.

Another way to increase axion abundance is to give the axion non-zero kinetic energy at the start of oscillations which can be used as an additional energy source. This kinetic misalignment mechanism has the initial axion field with non-zero angular momentum in field space, which then gets translated into a kinetic energy of the axion Goldstone boson~\cite{Co:2019jts}. This mechanism can lead to additional structure as well as GW production.

Finally, an additional energy source for the axion can enhance its abundance. One mechanism is 
where a heavy axion field decays into light axions through parametric resonance~\cite{Co:2017mop}. Another is axion ``friendship'' \cite{Cyncynates:2021xzw,Cyncynates:2022wlq}, in which two coupled ax\-io\-ns with nearby masses can exchange energy, with the axion with lower decay constant inheriting more energy from the one with higher decay constant. The latter mechanism does not appear to be applicable to the case of the QCD axion, but energy can be transferred by level crossing as the QCD axion mass turns on during the QCD transition~\cite{Cyncynates:2023esj}.

Suppressed axion abundance can be achieved through rearranging the axion's energy into other degrees of freedom, for example decays into relativistic matter that redshifts away quickly. One example is the axion decay to dark photons through parametric resonance~\cite{Agrawal:2017eqm}, which  reduces its abundance and potentially produces GWs and additional radiation.

\subsection{Axion Dark Matter Structure Formation}
\label{sec:structure}

While the QCD axion is indistinguishable from cold dark matter with current observations, there are interesting structure formation  differences due to the initial fluctuations of axions as well as due to their wavelike nature. For ultralight axions, beyond the standard QCD axion range, the wavelength becomes astrophysically relevant and has been explored in the context of the `fuzzy dark matter' regime~\cite{Hu:2000ke,Hui:2016ltb,Ferreira:2020fam,Hui:2021tkt}. The strongest current constraint with purely gravitational interactions of the axions is $m\gtrsim3\times10^{-19}~\ev$~\cite{Dalal:2022rmp} due to wave interference resulting in order-one density and thus gravitational potential fluctuations, causing heating of the stellar population dynamics. A range of other constraints from $m \gtrsim10^{-22}~\ev$ to $\gtrsim10^{-19}~\ev$ arises from suppressed power on short scales and altered stellar dynamics; for further details see e.g., refs.~\cite{Marsh:2015xka,Hui:2016ltb,Ferreira:2020fam,Hui:2021tkt}.

In many `more visible’ axion models described above which have with enhanced abundance compared to the traditional order-one axion misalignment abundance, the axion can have an enhanced matter power spectrum~\cite{Zurek:2006sy,Arvanitaki:2019rax,Cyncynates:2022wlq,Cyncynates:2021xzw}. This results in new compact structures which go by the names of  `oscillons', `solitons', `axion stars', and `minihalos'. If these structures survive into the present day they can dramatically change detection prospects compared to a smooth cold dark matter axion background; estimates result in an order-one fraction of dark matter in structure.

\begin{figure}[t]
\centering

\includegraphics[trim={0 0 0 0}, clip, width=\textwidth]{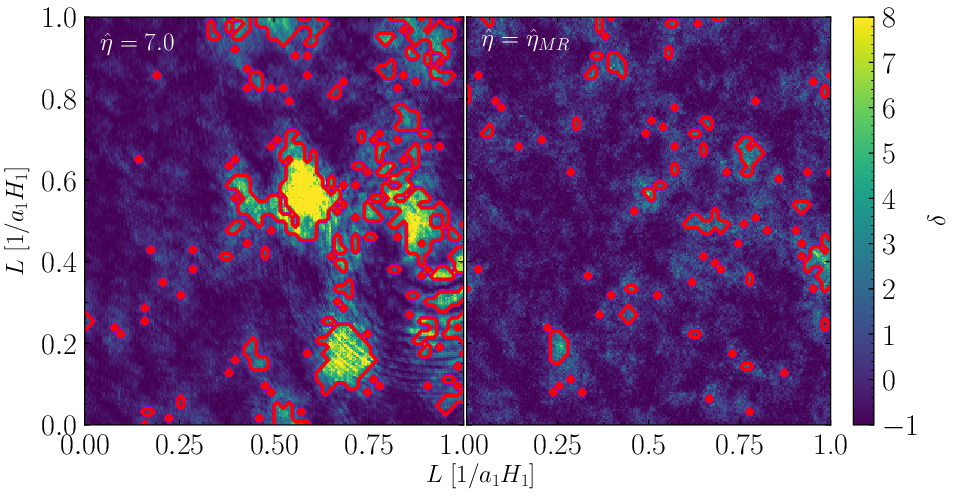}
    \caption{The overdensity of the axion field over the course of its evolution is shown as a 2D slice.  \textit{Left:} A portion of a 2D slice through the overdensity field at the end of the QCD crossover transition. Bright regions indicate large overdensities, and rings of relativistic radiation are visible as these overdensities decay. Clustered `minihalos' are outlined in red. \textit{Right:} As in left panel, but the overdensity field is shown at matter-radiation equality. At this time the large overdensities have dispersed, and the axion field is nonrelativistic. Figure from \citen{1906.00967}.
    }
    \label{fig:minihalos}
\end{figure}

Potentially even the standard QCD axion can impart unusual structure in our universe which may be observable. First, if PQ symmetry is broken after inflation, the resulting strings and domain walls can radiate axions which may contribute to a dark radiation component. In addition, as the universe passes through the QCD phase transition, large overdensities of axions form `miniclusters'~\cite{Hogan:1988mp,Kolb:1993zz,Kolb:1994fi,Kolb:1993hw,Fairbairn:2017dmf,Fairbairn:2017sil,1906.00967}, as seen in figure~\ref{fig:minihalos}. These large overdensities are thought to have mostly dispersed by the present day but may have an impact on structure formation. In the case of the QCD axion string formation scenario, it was recently pointed out that an order-one fraction of the dark matter today could be present in axion stars~\cite{Gorghetto:2024vnp}.

There are open questions regarding the fraction of axions in substructure and the resulting impact on searches and search strategies. It is important to establish how generically these structures form and survive until late times. They seem to be more prevalent in `more visible’ axion models, where the axion density is enhanced beyond that from misalignment. Their detection may thus require the development of new astrophysical searches as well as new laboratory search strategies.

\section{Astrophysics}
\markboth{ASTROPHYSICS}{ASTROPHYSICS}
\label{sec:astro}

Shortly after the axion was postulated, it was realized that
tight constraints on the axion parameter space could be extracted from observations of astrophysical objects.
A very early publication inaugurated bounds on Higgs and ax\-i\-ons from big bang nucleosynthesis,
microwave background and the abundances of
red giants \cite{dicus1978astrophysical}. 
Although by more recent standards those early bounds were not highly constraining,
perhaps surprising was that the most restrictive bound came from the ax\-i\-on's effects on red giant evolution.  Further important early astrophysical bounds used observables of our Sun's nuclear energy power emission~\cite{PhysRevD.36.2201} and the terrestrial neutrino flux from supernovae~\cite{Burrows:1988ah}.

These early bounds relied on the fact that the evolution of any star is throttled by the rate
at which thermal energy (carried away by photons, neutrinos, or, say, ax\-i\-ons) can escape.
Owing to the relatively short photon interaction length within stellar matter,
 stellar cooling is dominated by radiation from a relatively thin photosphere surface layer.
This applies as well to the core-collapse supernova SN1987A,
where core neutrinos are trapped and 
escape only from a relatively thin outer layer.
However, feebly-coupled ax\-i\-ons can have interaction lengths greater than
the size of the star and readily escape.
Since all the energy from axions produced anywhere within the bulk of the star escapes,
this radiation can be a significant cooling mechanism even if the ax\-i\-on is only rarely produced.
As a result of adding ax\-i\-on production  to normal stel\-lar-evo\-lu\-tion pro\-ces\-ses,
the star can either become colder or hotter, depending on the details.
The latter case arises when the star's gravitational potential energy becomes heat
when the additional axion cooling reduces the outward photon radiation pressure
thereby resulting in partial collapse.
Many of these, and other, early astrophysical bounds were detailed in
Georg Raffelt's seminal 1996
book ``Stars as Laboratories for Fundamental Physics''~\cite{Raffelt:1996wa}.

\begin{figure}[t]
\centering
\includegraphics[trim={0 0 0 0}, clip, width=1\textwidth]{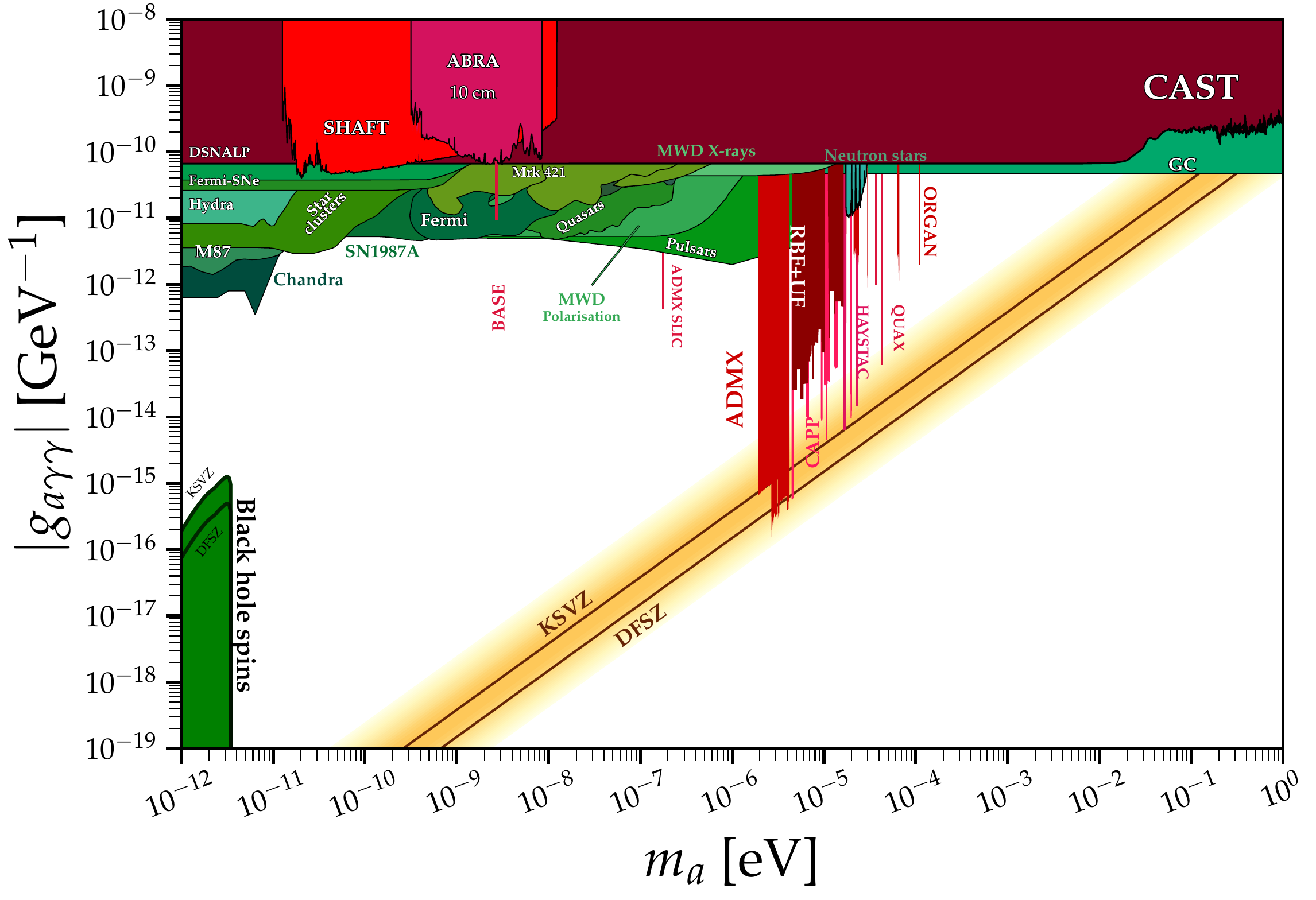}
    \caption{Astrophysical (green) and experimental (red) bounds on ax\-i\-on-pho\-ton coupling as a function of axion mass. See section~\ref{sec:astro} and section~\ref{sec:experiment} for details of the constraints.  The QCD ax\-i\-on benchmark parameters (KSVZ, DFSZ) are shown as a diagonal line from the lower left to the upper right.  Modified from Ciaran O'Hare~\cite{ciaran_o_hare_2020_3932430}. For further details, see e.g., the PDG
review \cite{Workman:2022ynf,ParticleDataGroup:2024cfk}.}    \label{fig:experimental_bounds}
\end{figure}

The general path for establishing astrophysical bounds is to test observations against the predictions of
stellar evolution with the ax\-i\-on added.
Of course, astrophysical predictions can have large uncertainties.
Hence, conservatism calls for reserving the claim of discovery to clear discrepancies between
stellar evolution models and observables. Since the early energy-loss constraints,  many ideas have  been developed to search for evidence of the axion in
astrophysical systems. New techniques and observables, from indirect observation of axions produced and converted to photons in stars and compact objects to superradiance clouds of axions produced by black hole spindown, have been realized and continue to extend the reach of astrophysical probes.
In this Section, we summarize the most important astrophysical searches and constraints which can impact the QCD axion parameter space, starting from less compact astrophysical objects such as the Sun and red giants (Sections~\ref{subsec:sun},~\ref{subsec:redgiants}), and increasing in compactness to white dwarves (Section~\ref{subsec:whitedwarfs}), neutron stars (Section~\ref{subsec:neutronstar}), and finally black holes (Section~\ref{sec:sr}). An overall summary of the bounds on the axion-photon coupling is shown in
figure~\ref{fig:experimental_bounds} and the QCD axion parameter space can be found in figure~\ref{fig:triptych}.

\subsection{The Sun}
\label{subsec:sun}

\subsubsection*{The Sun's luminosity:}~
At the center of the sun, ax\-i\-ons would be created in the Compton process
$e + \gamma \rightarrow e + a$.
For DFSZ axions,
the resulting axion total axion power density is about $10\,(m_a/\mathrm{eV})^2~\mathrm{erg /g /s}$.
Although our knowledge of the Sun's core thermodynamics from its
$^4$He abundance, luminosity and age is imperfect,
it is very unlikely the ax\-i\-ons at present could significantly contribute, readily bounding the DFSZ ax\-i\-on mass to less than
about an eV~\cite{PhysRevD.36.2201}.

Per the discussion above, an added axion energy transport channel would serve to
increase the temperature of the
Sun's core thereby increasing the Sun's neutrino flux measured on Earth.
However, no such neutrino excess is observed and such arguments rather robustly exclude ax\-i\-ons above around 1 eV,
where the production rate
in the core is too small, up to a few keV, the solar core temperature.
One could quibble with details, especially as they apply to stellar opacity,
and thus exactly where to place the mass bounds.

\subsubsection*{Direct detection of axions from the Sun:}~
 In addition to the cooling and stellar evolution based constraints described above, there are laboratory experiments on Earth that are capable of detecting a flux of ax\-i\-ons from the Sun and placing some of the strongest constraints across a wide range of parameter space of ax\-i\-on-like particles. The CAST experiment uses a large magnet to convert the ax\-i\-on flux from the Sun into x-ray photons, which would then be detected in an otherwise near-zero background environment
\cite{Anastassopoulos2017}
and has set constraints most recently in 2017 of
$g_{a\gamma\gamma}< 0.66 \times 10^{-10} \gev^{-1}$ at 95\% confidence for axion masses below $\sim10^{-2}$~eV (figure~\ref{fig:experimental_bounds}). Upgrades to BabyIAXO and IAXO could gain several or\-ders-of-mag\-ni\-tudes with larger ``target'' volumes, high\-er-fi\-eld magnets
and improved x-ray detectors \cite{universe8010037}.
The theoretical uncertainties of the solar flux are at the $\sim5$\% level and have been recently updated to include effects previously neglected, including the partial degeneracy of
electrons \cite{Hoof_2021}.

In addition to CAST, WIMP dark-mat\-ter experiments are also sensitive to ax\-i\-ons produced in the
hot solar core and then absorbed in, e.g., a Xenon atom through the ax\-i\-o\-elec\-tric effect~\cite{Derevianko:2010kz}.
Recent results from Xenon nT set joint constraints on the
$g_{a\gamma\gamma}< 5 \times 10^{-10} \gev^{-1}$ and  $g_{ae}< 2 \times 10^{-12} $ couplings;
the latter is a factor of a few weaker than the constraints from red giant cooling and white dwarf cooling (Sections~\ref{subsec:redgiants}, \ref{subsec:whitedwarfs})
but constitutes an important complementary
bound \cite{PhysRevLett.129.161805}.
	
While the bulk of stellar emission into ax\-i\-ons is into relativistic modes,
 heavy ax\-i\-ons with mass on the order of the solar temperature (keV) populate ``solar-basin'' bound orbits and can eventually
exceed the relativistic flux of ax\-i\-ons \cite{VanTilburg:2020jvl}.
The energy density can be detectable in
Earth-based laboratory experiments, allowing for direct or indirect detection of heavy ax\-i\-ons
without the assumption of a dark matter abundance.

\subsection{Red Giants and Horizontal Branch Stars}
\label{subsec:redgiants}

As main-se\-quence stars age, more and more hydrogen in the core is fused into $^4$He.
If the star has sufficient mass and enough $^4$He builds up, then $^4$He fusion processes in the core
proceed rapidly and the star evolves into a red giant.
These red giants are quite large, but with low temperature and luminosity.
The red giant $^4$He supply is ``burned'' quickly at a
core temperature around 10 keV; the lifetime of such red giants is short, only around 100 Myr.
After the $^4$He supply is exhausted, the red giant then continues evolving,
eventually becoming a compact object.

Ax\-i\-ons would be produced in the cores of red giants by a Pri\-ma\-kov-like process whereby
a real thermal photon scatters off a virtual photon near a nucleus,
the two photons thereby becoming an ax\-i\-on.
The ax\-i\-on readily transports energy out of the red giant,
thereby shortening the time the star spends in the red giant phase.

One observable used in placing ax\-i\-on bounds is the fraction of red giants in certain stellar populations,
the effect of ax\-i\-on emission would be to reduce this fraction.
This argument, applied to the red giant
populations in clusters gives a relatively robust upper ax\-i\-on-mass bound of a few
eV \cite{PhysRevD.36.2211}.

Should the ax\-i\-on be DFSZ (or otherwise have couplings to leptons),
then the ax\-i\-on-Comp\-ton process $e + \gamma \rightarrow e + a$ is important
and the dynamics becomes more complicated.
In slightly more detail,
the $^4$He core of the red giant is supported against collapse by electron degeneracy pressure.
As the star burns $^4$He, the star shrinks thereby converting gravitational potential energy
into heat and the star gets hotter.
Of course, there are still normal neutrino transport process at play,
cooling the core, but this neutrino cooling is relatively slow.
However, as is the case for neutrino production, the ax\-i\-on production rate in the core
is a high power of the core temperature,
and these ax\-i\-ons could effectively transport energy out of the core.
Hence, heavier and thereby more strong\-ly-coupled ax\-i\-ons remove more energy and depress the core temperature,
and as the ax\-i\-on mass increases, the $^4$He burning is increasingly quenched, thereby increasing the red giant lifetime.
Models of red giant dynamics, including the effect of DFSZ ax\-i\-ons, exclude such ax\-i\-ons with masses between around $10^{-2}$~eV where the production rate is too low, and the core temperature around
10 keV \cite{PhysRevLett.56.26}.
This, too, is a fairly robust bound but it, too, has uncertainties in the modeling leading to
``fuzziness'' on exactly where to place the mass bounds.

Current constraints using production of axions in red giants and horizontal branch stars use observations of old globular clusters (GC) in the galaxy and compare the numbers of horizontal branch stars to the number of red giant stars~\cite{Ayala:2014pea,Dolan:2022kul}. The ratio of these populations would be significantly affected by axion production in the cores of the starts. Most recent stellar evolution simulations quantify the uncertainties of stellar modeling and  constrain $g_{a\gamma\gamma}<0.47\times10^{-10} \mathrm{GeV}^{-1}$, which is comparable to CAST and places the leading constraint on axion-photon couplings at the high end of QCD axion parameter space, as shown in figure~\ref{fig:experimental_bounds}~\cite{Dolan:2022kul}.

\subsection{White Dwarfs}
\label{subsec:whitedwarfs}

White dwarfs are an interesting environment in which to search for ax\-i\-ons,
where the main focus of the search is in details of the white-dwarf luminosity function.
Ax\-i\-on emission, mainly from electron bremsstrahlung,
in the white dwarf atmosphere would speed up cooling and skew the luminosity function.
A somewhat recent study on white dwarfs in the the galactic disk
suggested that DFSZ ax\-i\-ons with masses
above around $10$~meV are
excluded \cite{Bertolami_2014}.
However the white-dwarf atmosphere model has considerable uncertainties and
the bounds may therefore be weak.
It is intriguing that an earlier analysis had
best fit to the luminosity function when a DFSZ ax\-i\-on
with mass of a few meV was added the white-dwarf
cooling~\cite{JIsern_2009}.
This analysis, while interesting, has not generated huge excitement as
the effect of this ax\-i\-on is small and the uncertainties are large.

Another probe for ax\-i\-on effects in white dwarfs is the pulsation period
of ZZ-Ceti white dwarfs.
The period slowing is related to the white dwarf cooling.
Analysis in this way of white dwarf cooling suggests again an
extra cooling channel consistent with a DFSZ ax\-i\-on as called for just
above~\cite{Corsico:2019nmr},
again with large uncertainties in the atmosphere model.
It is certainly interesting that two different analyses on white dwarfs
yield similar results.

On the other hand, such ax\-i\-ons emitted from magnetic white dwarfs would
convert into x-rays in the white-dwarf magnetosphere and those x-rays then detected in
orbiting x-ray observatories.
The Chandra x-ray observatory looked for such x-rays from the
magnetic white dwarf REJ0317-853 and did not detect any
excess~\cite{Dessert:2021bkv}.
This result is in conflict with the possible ax\-i\-on suggested from a
distortion of the white-dwarf luminosity function. See also~\cite{Dessert:2019sgw,Dessert:2022yqq,Ning:2024ozs} for constraints on axions with white dwarf physics and observations.

\subsection{Neutron Stars}
\label{subsec:neutronstar}

\subsubsection*{The Supernova Bound:}
The core-col\-lapse supernova SN1987A in the nearby LMC dwarf galaxy released
around $10^{53}$~ergs of gravitational potential energy
almost completely in neutrinos from a core temperature of tens of MeV.
The underground terrestrial detectors IMB and Kamiokande recorded between them
around 19 neutrinos over about a 10 second arrival time.
This was a milestone astrophysical observation,
confirming our gross understanding the supernova explosion with three neutrino flavors.
Ax\-i\-ons can be produced within the core via ax\-i\-on bremsstrahlung off of nucleons:
$N + N \rightarrow N + N + a$.
Neutrinos in the supernova core are trapped and the duration of the explosion
(and the temporal dispersion in the neutrino arrival times)
is therefore throttled by energy transport of neutrinos from the core.
The effect of adding in ax\-i\-on emission would be to add a new, efficient energy transport channel,
thereby shortening the duration of the explosion. 
Such arguments provide restrictive bounds,
excluding ax\-i\-on masses from around $10^{-3}$~eV, where the ax\-i\-on production is too feeble,
up to a few eV, where the ax\-i\-ons are trapped and otherwise approach the core
temperature \cite{PhysRevLett.60.1797}.

The supernova bound has has model uncertainties,
particularly the effect of nuclear properties on ax\-i\-on emission.
Hence, here, too, there is uncertainty in exactly where to place the mass bounds,
though it has been argued ax\-i\-ons with masses above around $10^{-2}$~eV are likely
forbidden \cite{Raffelt12345,Caputo:2024oqc}. Recent calculations exclude hadronic ax\-i\-on masses above $15-60$~meV, closing the hadronic (KSVZ) ax\-i\-on window at heavier masses~\cite{Chang:2018rso,Carenza:2019pxu}. Moreover, recent improvements in understanding pion populations in supernovae and mergers suggests that pion production may actually dominate ax\-i\-on bremsstrahlung from nuclei, through the process $\pi^- + p \rightarrow n + a$. If confirmed in simulations, pion emission further strengthens the ax\-i\-on mass bound from 1987A by a factor of about 2~\cite{Carenza:2020cis}.  Finally, observations of the 1987A remnant with ALMA~\cite{Carenza:2020cis,Page:2020gsx} and more recently JWST~\cite{fransson2024emission} has yielded firm evidence for the presence of the neutron star (NS) resulting from the supernova, putting these constraints on even more robust footing. In addition, studies of axion emission including higher-order chiral perturbation theory terms and finite density corrections  have shown that there is a model-independent constraint on the QCD axion, yielding $m_a<53^{+73}_{-13}$~meV~\cite{Springmann:2024ret}.

\subsubsection*{Neutron Star Cooling:}
In addition to the neutrino signal from Supernova 1987A, ax\-i\-ons can also contribute to the long-term cooling of NSs. Young, hot NSs have been used to place constraints;  Cassiopeia A from the seventeenth century supernova has been used to constrain ax\-i\-ons with decay constants comparable to those from SN1987A with some assumptions on the details of the NS interior~\cite{Leinson:2014ioa,1806.07151,Posselt:2018xaf}. Analysis of cooling data from NS HESS J1731-347 has yielded $f_a > 6.7\times 10^7$~GeV for KSVZ  and $f_a > 1.7\times 10^9$~GeV for DFSZ ax\-i\-ons~\cite{Beznogov:2018fda}, although there are some uncertainties on the nature of this compact object.  The strongest constraints come from a set of five NSs which are hundreds of thousands of years old with precise age measurements, resulting in an upper mass bound of $16$~meV for KSVZ models and $m_a > 10$ -- $30$~meV for DFSZ models (equivalent to couplings of $g_{an} < 1.3\times 10^{-9}, g_{ap} < 1.5 \times 10^{-9}$ at 95\% confidence)~\cite{Buschmann:2021juv}. 

\subsubsection*{Neutron Star Conversion:}

Ax\-i\-on dark matter passing through the magnetospheres of NSs can convert into radio photons in the large magnetic field.
Additionally, the conversion is resonant when the ax\-i\-ons pass through the region near the
NS where the mass of the ax\-i\-on matches the photon plasma
mass~\cite{Hook:2018iia} (see also \cite{Huang:2018lxq,Pshirkov:2007st}).
This allows NSs to act as targets for indirect ax\-i\-on detection with radio telescopes in the
range  $\sim 0.2$-$40\,\mu$eV.
 Improved treatment of the NS magnetosphere and environment \cite{Witte:2021arp}, as well as 3D modeling of the ax\-i\-on to photon conversion \cite{Millar:2021gzs} is expected to make the constraints more robust. Searches for signals from NSs have been performed in the data, with reach below the CAST bound but still far from the QCD ax\-i\-on expectation~\cite{Foster:2020pgt,Foster:2022fxn,Battye:2023oac}.
 
In addition to axion dark matter converting in the large magnetic fields of neutron stars, neutron stars themselves can source axions, either through the nuclear emission processes  described above, or through only the electromagnetic coupling in regions with large electric and magnetic fields (pulsar polar-cap cascades)~\cite{Noordhuis:2022ljw,Noordhuis:2023wid}. These axions can again be observed once converted through photons in magnetic fields. Observations of neutron stars have already placed constraints for emission in the neutron star~\cite{Foster:2022fxn} and in the polar caps~\cite{Noordhuis:2022ljw}. Interestingly, these observations lie in the radio band and in the event of a detection could provide motivation for haloscopes to zoom in on a particular region of axion mass and test this axion hypothesis.

\subsubsection*{Neutron Star Constraints on Exceptionally Light QCD Axions:}~
 Finite density effects change the ax\-i\-on potential. A lighter-than-expected ax\-i\-on can change properties of dense matter~\cite{Hook:2017psm,Balkin:2022qer}; if the ax\-i\-on has a tuned potential while still coupling to QCD, energy is minimized for field values $a\sim \pi f_a$ inside large, dense objects like NSs~\cite{Hook:2017psm}. 
In these exceptionally-light axion models, NSs source a non-trivial ax\-i\-on field which gives rise to axion radiation and  forces between the NSs for lighter axions~\cite{Huang:2018pbu}, and the lack of a NS crust for heavier axions~\cite{Balkin:2023xtr, Gomez-Banon:2024oux,Kumamoto:2024wjd}, which can be constrained with LIGO-Virgo-KAGRA (LVK) data~\cite{Huang:2018pbu,Zhang:2021mks} and NS crust observations~\cite{Gomez-Banon:2024oux,Kumamoto:2024wjd}, respectively. (See also~\cite{Hook:2017psm,Balkin:2022qer} for constraints on similar models from white dwarves.) Depending on the impact of a non-zero theta on nuclear interactions, these observables can reach to within 10\% of the standard QCD axion line~\cite{Kumamoto:2024wjd}; it remains to be seen whether high-density QCD at non-zero theta can be understood sufficiently well for dense matter to impact the profile of the minimal QCD axion. 

\subsection{Black Holes}
\label{sec:sr}

Black hole (BH) superradiance is a process through which rotating astrophysical BHs lose  energy and angular momentum and build up macroscopic, grav\-i\-ta\-tion\-al\-ly-bound states of ax\-i\-ons~\cite{Arvanitaki:2009fg,Arvanitaki:2010sy}. Note that superradiance is not limited to ax\-i\-ons; it can occur for any sufficiently weakly interacting boson \cite{Pani:2012bp,Baryakhtar:2017ngi,1704.04791,1705.01544,Baryakhtar:2020gao,Brito:2020lup}; see \cite{Brito:2015oca} for a detailed overview. For the purposes of this review we focus only on ax\-i\-ons. 

Superradiance leads to an exponential growth in the number of particles in the bound state, and can start from quantum fluctuations, thus not requiring a cosmological abundance of ax\-i\-ons. Superradiance occurs if the ax\-i\-on mass and bound state numbers satisfy the `superradiance condition' relative to the BH mass and spin: 
$\alpha \leq m\alpha_* / \left(2+\sqrt{1-\alpha_*^2}\right)$
, where $\alpha \equiv G_N M_{BH}  \mu $ is the dimensionless parameter governing the properties of growth rates and bound states, proportional to the ax\-i\-on mass $\mu$ and BH mass $M_{BH}$. The BH spin $a_*$ is a property of the Kerr metric and ranges from 0 (Schwartzchild BH) to 1 (extremal BH). The integer $m$ corresponds to the angular momentum along the rotation axis of the BH per ax\-i\-on in the bound state. States with lower total angular momentum have parametrically faster growth rates, so generally the lowest allowed $m$-value state dominates the evolution at a given time.

As the cloud grows, the BH spins down until the superradiance condition for the given level is saturated; most of the mass and angular momentum loss from the BH occurs in the last $e$-folding. The resulting bound state has on the order of $10^{77}$ ax\-i\-ons for stellar mass BH's, leading to new search strategies. We summarize three classes of signatures: BH spin-down; gravitational wave (GW) emission; and gravitational impact of the cloud on dynamics near the BH, and then comment on effects of non-gra\-v\-i\-ta\-tional ax\-i\-on couplings.

\subsubsection*{Black Hole Spin Constraints:}~
As superradiance causes BHs to spin down on a characteristic timescale set by the ax\-i\-on mass, high\-ly-spin\-ning BHs can be used to constrain the parameter space of light, weak\-ly-coupled ax\-i\-ons~\cite{Arvanitaki:2010sy,Arvanitaki:2014wva}. Measurements of BH spins and masses have been performed generally in binary systems, either by measuring the innermost stable circular orbit by observing the accretion disk from the companion star or by fitting gravitational waveforms to identify the BH spin. Several BHs in X-ray binaries have been observed to have very high spins, which  can be used to constrain ax\-i\-on parameter space as shown in figure~\ref{fig:experimental_bounds}.  Conservative constraints on ax\-i\-on parameter space from BH spin measurements are shown \cite{Baryakhtar:2020gao} with the standard conversion between $1/f_a$ and $g_{a\gamma\gamma}$ for the KSVZ and DFSZ models, constraining QCD axion masses to be above $\sim 6\times10^{-12}$ eV from measurements of Cygnus X-1 and slightly weaker constraints from other black holes. More stringent constraints which combine data from multiple BH measurements with specific assumptions about systematics and error covariance increase the reach to higher ax\-i\-on masses~\cite{Mehta:2020kwu}; a full statistical analysis with M33 gives $m_a\gtrsim 3\times10^{-12}$ eV~\cite{Hoof:2024quk}.  

	\begin{figure}[th!]
	\begin{centering}
		\includegraphics[width=0.8\textwidth]{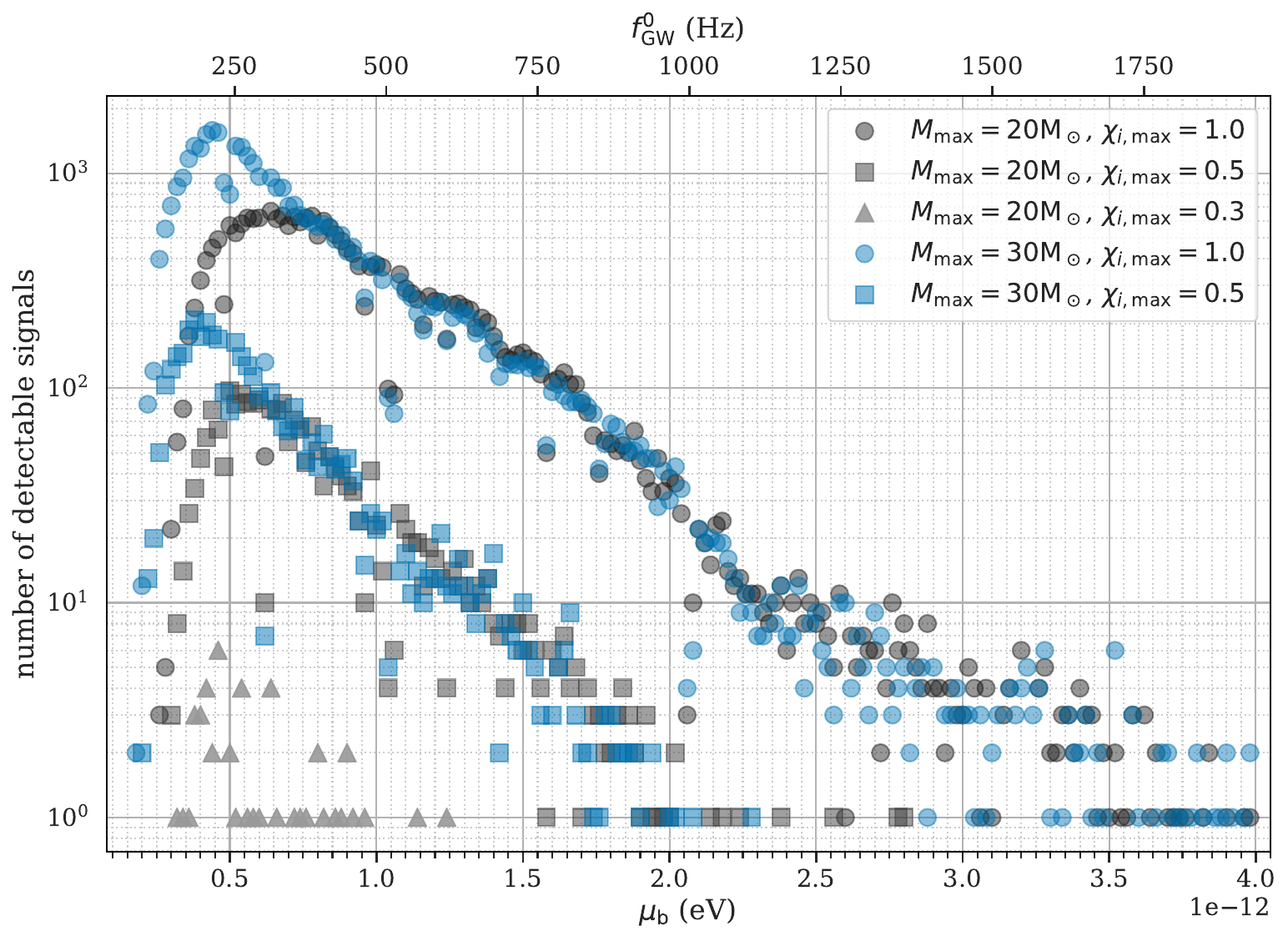}
		\caption{Number of gravity wave signals with amplitudes above the detectability level of recent all-sky searches for continuous GWs \cite{FalconO1,Dergachev:2019oyu, Palomba:2019vxe}, for bosons in the mass range $1\times10^{-13}$~eV  to $4\times10^{-12}$~eV. We consider the standard (black circles) and heavy (blue circles) BH populations, as well as the moderate (gray squares), heavy moderate (blue squares) and pessimistic (gray triangles) spin populations. The fluctuations are due to  stationary lines and other artifacts in the detectors, which decrease the CW search   sensitivity in nearby frequency bins. Figure from \cite{Zhu:2020tht}.}
	\label{fig:srsignals}
	\end{centering}
\end{figure}

It is also possible to use statistical analyses of the BHs merging in LIGO data to constrain---or search for signatures of---superradiant spin down~\cite{Arvanitaki:2016qwi}. Gaining reliable evidence for the ax\-i\-on hypothesis likely requires hundreds of events~\cite{Arvanitaki:2016qwi,Ng:2019jsx}; on the other hand,  a factor of 2 exclusion in ax\-i\-on mass, $1.3\times 10^{-13}$~eV-$2.7\times 10^{-13}$~eV (assuming sufficiently weak self-in\-ter\-ac\-tions and sufficiently long BH merger times) has been placed, driven largely by two systems, GW190412 and GW190517, which were observed to have non-zero initial BH spin at high significance~\cite{Ng:2020ruv}.

Measurements of supermassive BHs can probe ultralight ax\-i\-ons~\cite{Arvanitaki:2014wva,Arvanitaki:2016qwi,Stott:2018opm,Cardoso:2018tly,Mehta:2020kwu, Du:2022trq}. However, the spin measurements of supermassive BHs are more uncertain, and superradiant evolution times become comparable to astrophysical timescales such as accretion and mergers, so bounds have to be interpreted with extreme care.
 
\subsubsection*{Searches with Gravitational Waves:}

The ax\-i\-on bound states around the BH constitute an enormous energy density with both angular and time dependence. 
In general relativity, these clouds then necessary source GWs. Specifically, the components of
the stress energy tensor with quadrupole or higher angular dependence, and
non-zero time dependence, source GWs. At the particle level, the high-fre\-quen\-cy
terms correspond to annihilations
of two bound-state ax\-i\-ons to a single graviton with frequency of order twice the ax\-i\-on mass, and are always present, while low frequency terms correspond to
transitions from a higher-energy level to a more deeply bound one, and are present when multiple levels are simultaneously populated~\cite{Arvanitaki:2010sy, Arvanitaki:2014wva}. The latter are thus more rare except in the case where ax\-i\-ons self-interact to a significant extent~\cite{Arvanitaki:2014wva,Baryakhtar:2020gao}.

These GWs lead to energy loss from the ax\-i\-on cloud, on timescales parametrically slower than the superradiance rate. Thus the peak `annihilation' signal takes place once the cloud has finished growing through superradiance and can last for a period ranging from days to millions of years, depending on the ax\-i\-on and BH masses.  The signals from individual BHs can be observed in blind \cite{Arvanitaki:2014wva,Arvanitaki:2016qwi} or directed \cite{Arvanitaki:2016qwi,Ghosh:2018gaw} continuous wave searches in LVK data~\cite{Riles:2022wwz}. Alternatively, a large population of BHs which are all emitting GWs can manifest as excess stochastic noise in a frequency band around twice the ax\-i\-on mass \cite{Brito:2017wnc,Brito:2017zvb,Tsukada:2018mbp}. Both the blind and stochastic signals have a strong dependence on the properties of the BH population (in the Milky Way and beyond), in particular the formation rate of BHs, and their mass and especially spin distributions at formation~\cite{Zhu:2020tht}. With current data and standard BH population assumptions, between several and several hundred monochromatic continuous wave signals for ax\-i\-ons of around $10^{-13}$~eV in mass would be expected in LVK data and all searches have been null to date \cite{Palomba:2019vxe,Zhu:2020tht,LIGOScientific:2021jlr}; however any exclusion in ax\-i\-on parameter space is conditional on the unmeasured properties of Milky Way BHs.

Directed searches, following up a merger with a newly born BH, is the ``golden observation'' for exclusion or detection of a new particle. The BH formation time and properties being well-measured, the superradiance waveform can be calculated as a function of ax\-i\-on mass. Most likely, next-genera\-tion GW observatories will be required for such searches for ax\-i\-ons~\cite{Arvanitaki:2016qwi,Ghosh:2018gaw,Isi:2018pzk}, but present detector sensitivity is already promising for vector particles~\cite{Jones:2023fzz}.

In addition to  the currently ongoing BH spin and GW measurements, future observations may be able to shed light on the presence of the cloud around BHs, such as the modification of BH inspirals \cite{Baumann:2018vus,Hannuksela:2018izj,Baumann:2021fkf,Baumann:2022pkl}, back reaction of the cloud to create `floating orbits' \cite{Cardoso:2011xi,Zhang:2018kib}, and the imprint of superradiance on hierarchical BH mergers \cite{Zhang:2019eid,Payne:2021ahy}.

\subsubsection*{Non-Gra\-vi\-ta\-tional Interactions:}
Superradiance is a fundamentally gravitational process and thus does not depend on any interactions of the particle other than its mass and spin. However, the large field amplitudes that are formed in the final stages of superradiant growth can impede or otherwise change the dynamics of superradiance in the presence of non-grav\-i\-ta\-tional couplings~\cite{Arvanitaki:2010sy}.  In the case of ax\-i\-ons, the most relevant is the self-in\-ter\-ac\-tion, at leading order a quartic (Section~\ref{sec:intro}). The self-in\-ter\-ac\-tion leads to new dynamics in the cloud, enabling the ax\-i\-on field to source new bound angular momentum levels as well as non-relativistic and relativistic ax\-i\-on waves; the former could  be seen in future earth-based laboratory experiments searching for ax\-i\-on DM~\cite{Baryakhtar:2020gao}. The self-in\-ter\-ac\-tions also limit the maximum field value of the cloud~\cite{Yoshino:2012kn,Gruzinov_2016,Baryakhtar:2020gao,Omiya:2022gwu}, which slows down spin extraction from the BH and thus relaxes the bounds for lower-$f_a$ axions~\cite{Baryakhtar:2020gao} and reduces GW signals~\cite{Baryakhtar:2020gao,Collaviti:2024mvh}. There has been uncertainty in the literature regarding the existence of a `bosenova', a violent collapse of the cloud under self-in\-ter\-ac\-tions; it was recently demonstrated that for small enough ax\-i\-on masses relative to BH size, the bosenova does not occur \cite{Baryakhtar:2020gao,Omiya:2022gwu}. Further work is needed to establish dynamics of higher angular momentum levels and relativistic clouds, which introduce additional uncertainties but could extend the reach of BH superradiance to heavier QCD ax\-i\-ons by a factor of about 2 \cite{Baryakhtar:2020gao, Omiya:2024xlz, Hoof:2024quk}.

The axion coupling to photons can in principle lead to parametric resonance decays of the cloud to photons, although the finite size of the cloud~\cite{Hertzberg:2018lmt} and the presence of interstellar plasma~\cite{Sen:2018cjt}  impede this process in realistic situations~\cite{Baryakhtar:2020gao}. The large ax\-i\-on field can also act as a type of magnetic field acting on spins in the BH vicinity; however, the maximal values do not exceed a fraction of a gauss, which is negligible compared to, e.g., accretion disk B fields~\cite{Arvanitaki:2010sy}. For photon couplings enhanced by several orders of magnitude compared to the KSVZ and DFSZ expectations, a complete analysis in a realistic BH environment has yet to be performed. 	

\section{Laboratory Experiments}
\markboth{LABORATORY EXPERIMENTS}{LABORATORY EXPERIMENTS}

\label{sec:experiment}

As discussed throughout this review, the coupling of the axion to normal
matter and radiation is extraordinarily feeble.
It would be perhaps therefore surprising that laboratory
axion searches would be promising for direct detection
of the QCD dark-mat\-ter axion.
Nonetheless, laboratory experiments have been demonstrated
to be exquisitely sensitive and are a promising
path to discovery. 

We first place the axion scales relevant for experiments in context in Section~\ref{subsec:lengthscales}. 
Before we discuss the more recent developments in axion experiment,
it is worthwhile to review the outcome of the early terrestrial accelerator and reactor experiments launched in the decade or two after the axion was postulated (Section ~\ref{sec:earlybounds}).
Early on, it was appreciated that bounds on ax\-i\-ons and other hypothetical
particles could be established by their observable effects on stellar evolution  and were many orders of magnitude more restrictive than these searches (see Section~\ref{sec:astro}).
It was not until the microwave cavity concept that attention returned to terrestrial QCD ax\-i\-on dark-mat\-ter experiments. This line of inquiry remains very active today  (Section ~\ref{subsec:haloscope}) and is being continuously developed as described in Section~\ref{sec:futurecavity}. We summarize other approaches to detect the dark matter axion through the electromagnetic coupling in Section~\ref{subsec:variants} and outline ongoing and proposed searches for axion-nucleon couplings in Section~\ref{subsec:nucleon}.

\subsection{Relevant Length Scales}
\label{subsec:lengthscales}

\begin{figure}[th!]
\centering
\includegraphics[trim={2cm 0 1cm 0}, clip, width=1.0\textwidth]{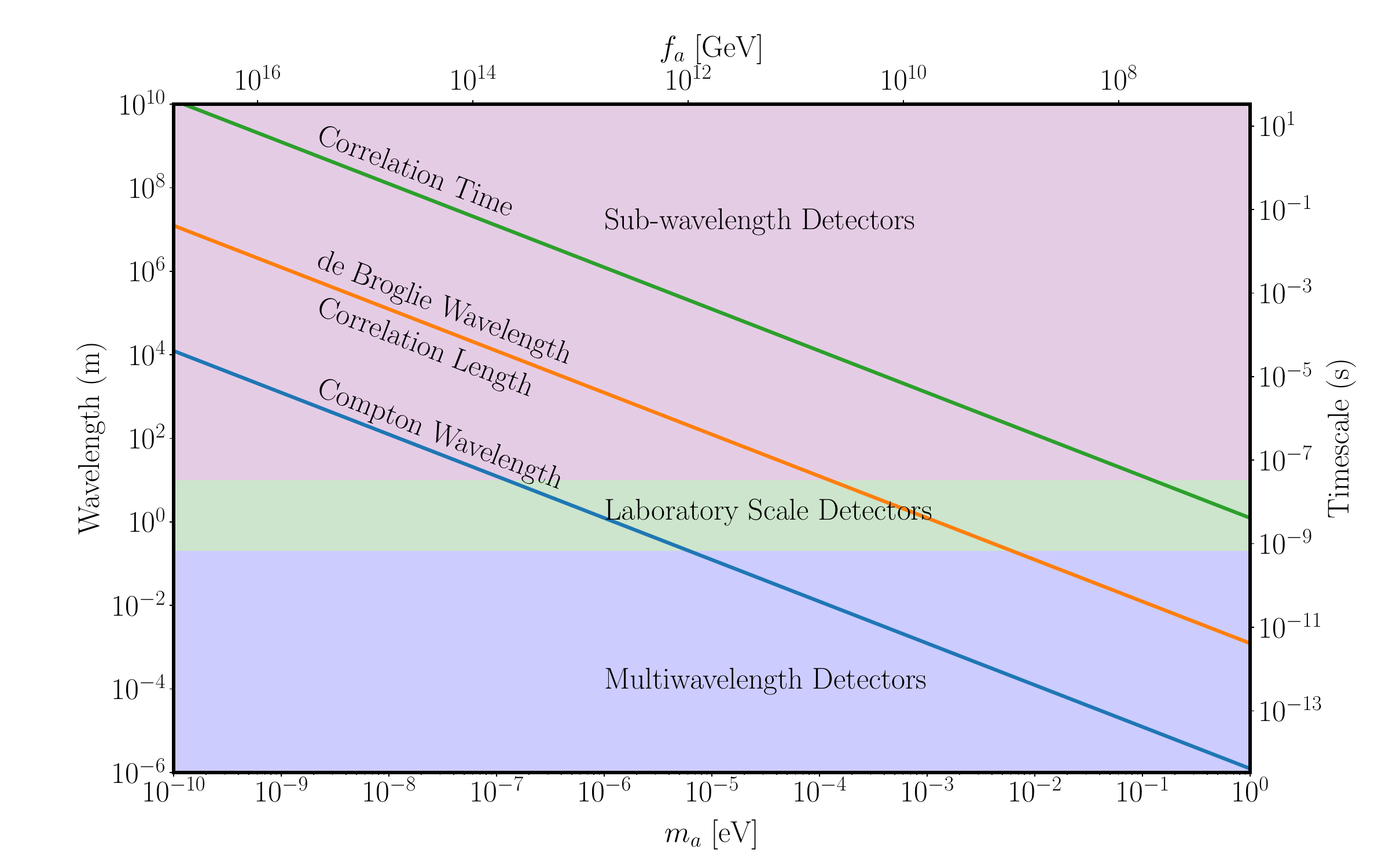}
    \caption{Characteristic lengths and times for virialized axion dark matter as a function of mass.  Dark matter is non-relativistic so the scaling of Compton and de Broglie wavelength is a trivial scaling with axion mass and velocity, but the correlation scales are more dependent on assumptions about the dark matter kinematics as described in the main text.  }
    \label{fig:wavelength}
\end{figure}

For the small masses left unexplored by early laboratory and astrophysics searches, the axion is distinct from most subatomic particles in that its relevant length scales can be much larger than those of the detector.  Axions, depending on their mass, have characteristic length scales that are somewhat smaller than, comparable to, or larger than the experimental apparatus.  This leads to a wide variety of techniques, each applicable to only a certain range of masses.  The relevant length scales to consider are the Compton wavelength, which is the wavelength of photons created from axion conversion, the de Broglie wavelength, which is the length scale of the axion matter wave, and the correlation length and time, which are the length and time scales over which the phase of the axion field is coherent.

Lower bounds on the correlation length and time can be estimated from the maximum velocity dispersion axions can have in order to make up a virialized dark matter Milky Way halo. For the Maxwellian distribution of a fully virialized halo, all the length scales -- the mean and the most probable de Broglie wavelengths as well as the correlation length -- are set by the virial velocity $v_0$ in the halo and are within tens of percent of one another. The correlation time is longer by another factor of $c/v_0$. In principle, axions may be much colder with much longer correlation lengths and times.  This occurs for example for subcomponents of dark matter that are not virialized such as streams, or axions in self-bound structures such as miniclusters.   The relationship between these length scales and the laboratory length scale is shown in figure~\ref{fig:wavelength}.  In the event of a detection, the information from these correlations can be used to extract information about the dark matter halo (e.g., ref.~\cite{Foster:2017hbq}). Existing ax\-i\-on-like particle bounds from astrophysics and laboratory experiments are shown in figure~\ref{fig:experimental_bounds}, where it is notable that the most sensitive laboratory experiments correspond to the mass range where the Compton wavelength is roughly at the laboratory scale.

\subsection{Early Searches, Limits and Bounds}
\label{sec:earlybounds}

Recall that early searches for axions assumed the PQ scale to be near the electroweak scale,
and such axions would be relatively strongly coupled to normal matter and radiation and would have been produced
in abundance and detected in reactor and accelerator experiments.
The negative searches imply the axion scale $f_a$ is considerably greater than the weak scale.
Following, we give an updated summary of these early searches, largely adapted from the
reviews~\cite{KIM19871,CHENG19881,ROSENBERG20001}.

The experimental searches into the early 1980's were broadly divided into terrestrial experiments
(reac\-tor-based, ac\-cel\-er\-a\-tor-based, etc.) and astrophysical bounds (counting horizontal branch stars, e.g.).
Although the early terrestrial experimental bounds played the important role of decoupling the axion scale from the weak scale,
the feebleness of the `invisible' axion couplings now largely render such early search methods (reactors, beam dumps, etc.) obsolete;
they lack the required sensitivity to QCD axions in the most plausible mass range
by orders of magnitude and other terrestrial technologies have overtaken them. Similar and new concepts are still active today in the search for `heavy axions' (see, e.g., refs.~\cite{NOMAD:2000usb,Jaeckel:2015jla,Knapen:2016moh,CMS:2018erd,Aloni:2019ruo,ATLAS:2020hii,NA64:2020qwq,Capozzi:2023ffu,BESIII:2024hdv}).

\subsubsection*{Reactor experiments}
\begin{figure}[ht]
\centering
\includegraphics[width=0.65\textwidth]{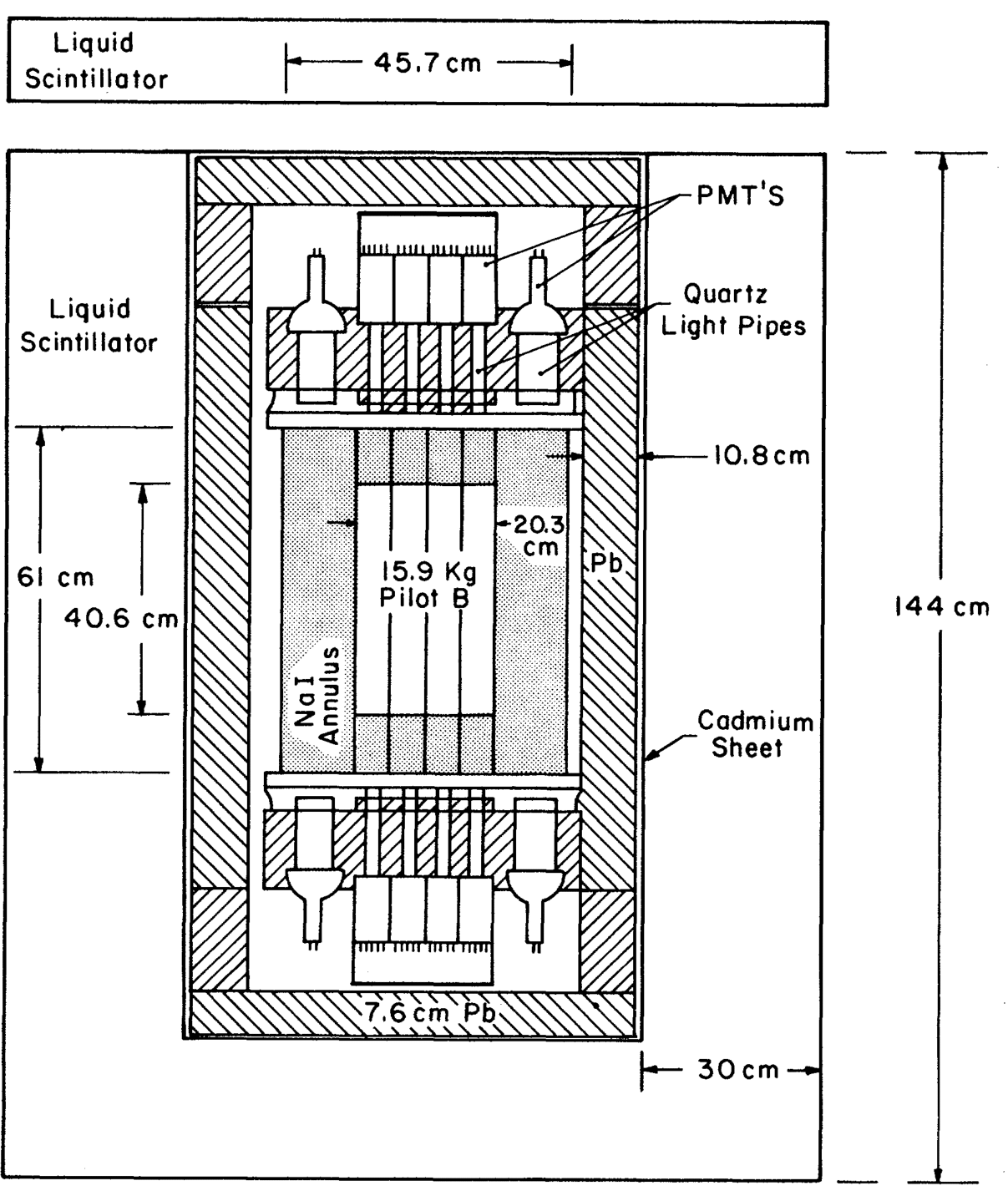}
\caption{Sketch of the Rei\-nes-Gurr-So\-bel $\bar{\nu}_e$ scattering experiment at
the Savannah River power reactor. The main detector consists of a plastic scintillator target.
An axion of appropriate mass and couplings would have resulted in a substantially
increased electromagnetic count rate.
From \citen{PhysRevLett.37.315}.}
\label{reines-gurr-sobel-experiment-diagram}
\end{figure}

The core of a nuclear reactor is the source of a huge flux of $\bar{\nu}_e$'s.
Although the yields can vary considerably, approximately each $\bar{\nu}_e$ is accompanied by a prompt $\gamma$;
this $\gamma$ production is dominated by E1 transitions with a small component of M1 transitions,
where most of the M1 $\gamma$'s arise from isovector transitions.
However, should there be an axion of appropriate mass and coupling, there would as well be a pseudoscalar contribution
increasing the M1 rate via direct axion emission.
These fi\-nal-state axions could escape the reactor core and subsequently decay into two photons,
which would then be detected by nuclear scintillation detectors.
An alternate detection process is for the emitted axion to convert into a photon through the Compton process
$a + e^- \rightarrow \gamma + e^-$ where the outgoing $\gamma$ is detected.
In one such experiment\cite{PhysRevLett.37.315},
(sketched in figure~\ref{reines-gurr-sobel-experiment-diagram}) NaI scintillation detectors were positioned
approximately 1 m from the Savannah River reactor core.
Energy depositions of greater than 1.5 MeV were counted, yielding a back\-ground-sub\-trac\-ted counting rate of
$-160 \pm 260$ counts/day, whereas
the count rate from weak-scale axions would result in rate
of over 1000 counts/day.
With no excess counts observed, this experiment and other reactor experiments began to strain the weak-scale axion hypothesis.

Other re\-ac\-tor-ex\-per\-i\-ment results were then reinterpreted in the context of axion emission.
An even-ear\-li\-er experiment\cite{PhysRevLett.32.180}
searched for weak neutral current neu\-tri\-no-in\-duced disintegrations
of deuterium via the process $\bar{\nu}_e + d \rightarrow \bar{\nu}_e + n +p$.
An axion with energy in the MeV range would have added approximately 1000 counts/day from the process of
induced disintegration from nucleon scattering to the observed back\-ground-sub\-trac\-ted count rate
$-3 \pm 7$ counts/day. Again, a weak-scale axion would have been readily observable.

\subsubsection*{Nuclear de-excitations}
A nucleus in an excited state can de-excite through axion emission, with the emitted ax\-i\-on-nu\-cle\-us system
carrying away spin-par\-i\-ty $J^{P}=0^-, 1^+, 2^-, 3^+ ...$ etc.
Specifically, for $J^P = 1^+$, the isoscalar and isovector transition rate has been estimated for several isotopes where the
de-excitation has been measured and is well understood\cite{PhysRevD.18.1607}.
For $^{7}$Li, $^{97}$Nb and $^{137}$Ba, the observed de\-ex\-ci\-ta\-tion rates are considerably less than that from a weak-scale axion.
These early low-en\-er\-gy nuclear experiments showed promise, but,
since the time of these early bounds, we know of no new high\-er-sen\-si\-ti\-vi\-ty results for nuclear de-excitations in the context of axions.

\begin{figure}[ht]
\centering
\includegraphics[trim={5cm 16cm 5cm 15cm}, clip, width=0.85\textwidth]{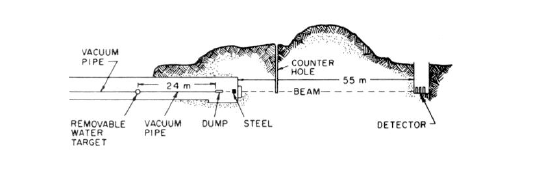}
\vspace{.4cm}
\includegraphics[trim={1cm 21.5cm 1cm 7cm}, clip, width=0.85\textwidth]{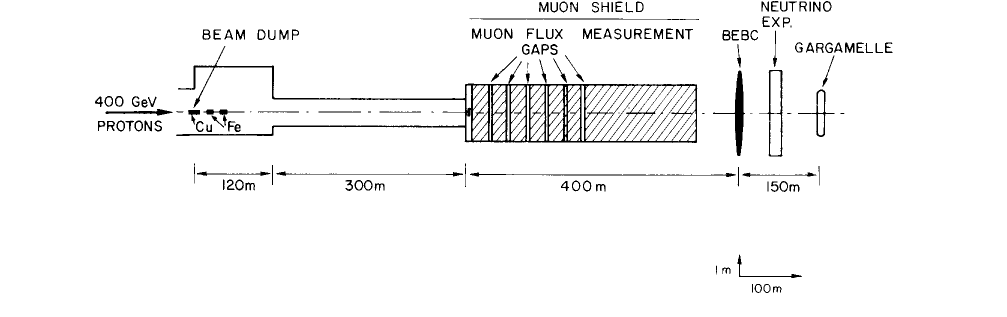}
\caption{\textit{Top:} Sketch of the Donnelly et al. beam-dump experiment at SLAC end-station A.
From \citen{PhysRevD.18.1607}. \textit{Bottom:} Sketch of the BEBC beam-dump experiment at the CERN SPS.
From \citen{BOSETTI1978143}.}
\label{donnelly-etal-experiment-sketch}
\end{figure}

\subsubsection*{Beam dumps}
Beam dumps have a long history in searching for penetrating particles.
The dump mass shields the detector from the source and allows a much-re\-duced background with a consequent increase in sensitivity.
In one type of experiment, axions in such searches would be produced in the source by directing a particle beam into a target.
The resulting axions then penetrate an earth berm (the `dump' or `shield').
Axions or their decay products then interact with a detector.
Early on, SLAC had a major role in these beam-dump axion experiments.
In an early search \cite{ROTHENBERG0,ROTHENBERG1,ROTHENBERG2}, reinterpreted in the context of axions \cite{KIM198155},
electrons from the linac were directed onto a target.
The beam-dump geometry is shown in figure~\ref{donnelly-etal-experiment-sketch}~(upper).
In the target, axions would be produced via the axion bremsstrahlung process off the nuclear electromagnetic field.
These axions would then pass through a 55 m earth berm and then decay into muon pairs, either in flight
or in the berm near the exit.
The resulting two-muon event signature is distinctive, and after analysis no muon pairs were detected.

Several CERN groups also searched for axions in beam dumps~\cite{ALIBRAN1978134,HANSL1978139,BOSETTI1978143}. 
The axions in this case were produced by directing 400 GeV SPS-pro\-duced protons onto a target.
Axions, produced by a nuclear bremsstrahlung process in the nuclear field, would then pass through
an earth berm and decay into electron, muon or possibly neutrino pairs, either in flight or near the exit of the berm.
The detectors were placed approximately 1 km from the target.
A small number of observed lepton pairs were attributed to neu\-tral-cur\-rent dilepton production.
However, weak-scale axions would have resulted in a much larger number of pairs  than that expected from the neutral current background.
The configuration of one such CERN experiment is shown in figure~\ref{donnelly-etal-experiment-sketch}~(lower).

Interest in beam dump experiments increased with the reports of anomalous $e^+e^-$ pairs produced in heavy ion
fixed-tar\-get collisions at GSI (see, e.g., ref.~\cite{PhysRevLett.56.444}).
One interpretation of these pairs was that they arose from the decay of a 1.8 MeV/c$^2$ axion-like particle
produced in the collision.
In retrospect, the peaks were likely spurious (see, e.g., ref.~\cite{doi:10.1126/science.275.5297.148}).
Later SLAC experiments were then targeted at detecting these possible GSI axions.
SLAC experiment E-141 directed 9 GeV electrons from the linac onto a combined tungsten target
and dump~\cite{PhysRevLett.59.755}.
Again, bremsstrahlung axions would be emitted in the collisions and escape from the exit of the dump.
The axions would decay in flight and the decay products would be detected by the venerable SLAC 8 GeV spectrometer.
Backgrounds from hadrons and muons were identified with a hydrogen Cerenkov detector and lead-glass calorimeter.
By the time those data were analyzed, the `onia' experiments, described below, had already excluded a weak-scale axion
in the accessible mass range.
So, the E-141 focus was on an unusual `GSI-excess' axion that would preferentially decay into $e^+e^-$ pairs
with lifetime appropriate for the distance between the exit of the dump and the detector.
Such `leptophilic' axions do occasionally appear in the
literature (ref.~\cite{PhysRevD.103.035028} is one of many such examples).

\begin{figure}[!htb]
\centering
\includegraphics[width=0.5\textwidth]{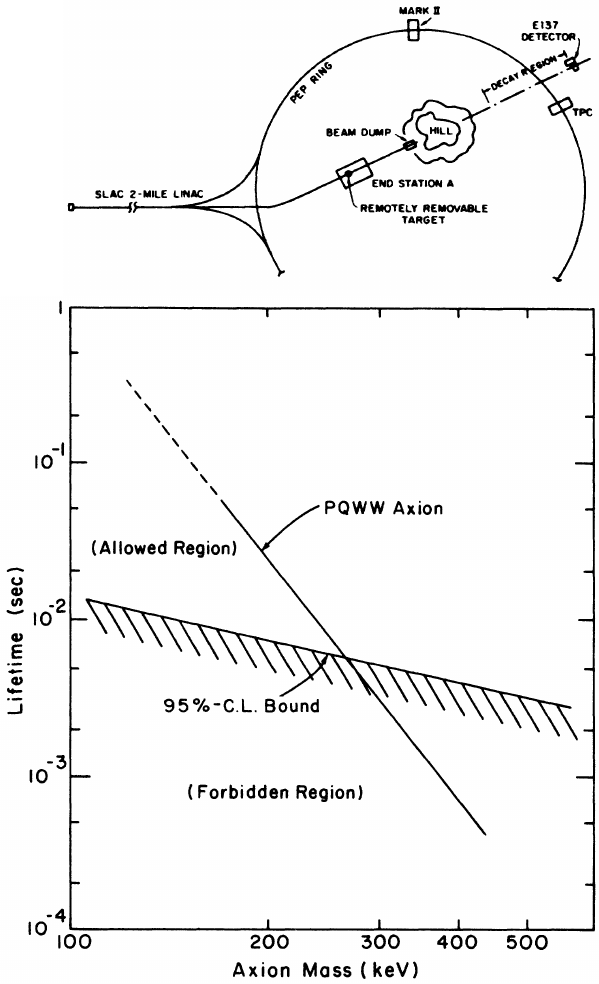}
\caption{(upper) Sketch of the E-137 beam-dump experiment at SLAC.
(lower) E-137 limit on the Pec\-cei-Quinn-Wein\-berg-Wil\-czek axion.
This weak-scale axion is excluded below the few 100 keV axion mass range.
From \citen{PhysRevD.38.3375}.}
\label{Eonethirtyseven}
\end{figure}

Perhaps the most sensitive of these early beam-dump experiments for the purposes of this
review was SLAC E-137 \cite{PhysRevD.38.3375},
schematically shown in figure~\ref{Eonethirtyseven} (upper).
Here, 20~GeV electrons from the linac were directed at an Al/H$_2$O target and axions produced in the target
would then penetrate a 179 m earth berm.
On the exit side of the berm was the detector consisting of an 8 radiation length electromagnetic shower calorimeter,
especially sensitive to electron or photon pairs.
Axions with masses greater than about a few 100 keV were thereby excluded, with
the limit shown in figure~\ref{Eonethirtyseven} (lower).

\subsubsection*{`Onia' decays}
The 1$^-$ heavy quark`onia' states (the J/$\Psi$ and B states) can decay into $a + \gamma$.
If the axion lives sufficiently long, it escapes and the experimental signature is a distinctive single photon.
The `onia' detectors consist of highly segmented electromagnetic calorimeters,
thus yielding an almost back\-ground-free sin\-gle-pho\-ton signature for axion production.
One can also imagine a short-life\-time axion decaying in flight into electromagnetic products before entering the calorimeter,
also yielding an almost back\-ground-free signature.
A series of collaborations: CUSB\cite{PhysRevD.26.717}
and CLEO\cite{PhysRevD.27.1665} at
the CESR $e{^+}e{^-}$ storage ring at Cornell,
 Crystal Ball~\cite{PhysRevLett.48.903}
at the SPEAR $e{^+}e{^-}$ storage ring at SLAC,
and LENA~\cite{Niczyporuk1983}
at the DORIS $e{^+}e{^-}$ storage ring at DESY,
looked for the single photon event topology and at very high confidence excluded the weak-scale axion.
Somewhat later CUSB and CLEO studies excluded shor\-ter-life\-time weak-scale axions with 
a single photon plus two electromagnetic (e or $\gamma$)
particle final state~\cite{PhysRevLett.56.2672,PhysRevLett.56.2676}.

\subsection{Microwave Cavity Haloscopes}
\label{subsec:haloscope}

\begin{figure}[th]
\centering
\includegraphics[width=0.55\textwidth]{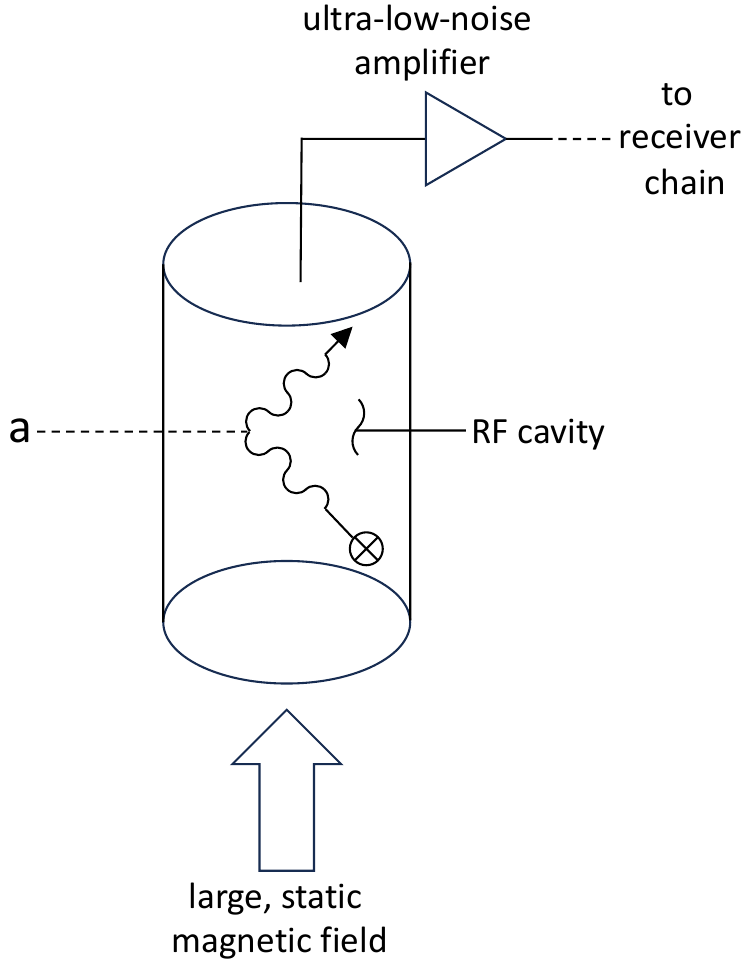}
\caption{The Si\-ki\-vie-con\-cept haloscope axion detector. Cosmic halo axions scatter off a large, axial,
static magnetic field and thereby convert into a microwave photons inside a tunable high-Q RF cavity.
The experimental axion signature is
very slight excess cavity electromagnetic power at the frequency of the cavity TM$_{010}$ mode when that mode is
tuned to the axion mass.}
\label{fig:sikivieconcepthaloscope}
\end{figure}

Given these experimental constraints, as well as the astrophysical bounds described in the previous section, it had been assumed that the extraordinarily feeble
couplings of the QCD dark-mat\-ter axion (the `invisible axion') to
normal matter and radiation
would render those axions impossible to detect in experiments, and progress in experimental searches slowed from that point on.  
However, in 1983 Pierre Sikivie conceptualized the microwave cavity haloscope whereby invisible axions could be detected~\cite{PhysRevLett.51.1415}
(also see refs.~\cite{PhysRevLett.52.695.2,Krauss:1985ub}).
The axions to be detected this way are ``cosmic axions'', nearby Milky Way halo
dark-mat\-ter axions. Since these axions are
already in great abundance in the vicinity of a terrestrial laboratory (with a prodigious axion number
density of $\sim 10^{15}/$cm$^3$ at $\mu$eV masses), the tiny factor of the axion coupling
to normal matter and radiation need not be expended in producing them.
The haloscope includes a large right-cir\-cu\-lar-cyl\-in\-der high-quality factor RF cavity
threaded axially by a large static magnetic field, see figure~\ref{fig:sikivieconcepthaloscope}.

From a microscopic physics perspective, a single nearby halo axion
scatters off of a ``static'' virtual
photon in the magnetic field, thereby converting into a single real photon
populating the TM$_{010}$ cavity mode.
The axion field at the relevant mass is coherent over the
cavity volume and, for reasonable magnet and cavity parameters, the
conversion rate is in the neighborhood of 100 Hz or less for the more-weak\-ly
coupled, but highly compelling DFSZ axion. The corresponding signal power
is in the yoctoWatt range or less. This ax\-i\-on-con\-ver\-sion process is,
in spirit, similar to the inverse of Primakov scattering of neutral pions off a nuclear
electromagnetic potential, thereby converting neutral pions into photons
in the nuclear electromagnetic field.\footnote{This analogy is close since
axions and neutral pions can mix.} The high
cavity quality factor $Q$ enhances the conversion rate; one way to picture this $Q$ enhancement is
to recall a high quality factor increases the effective density of states near
cavity resonant frequency, thereby increasing the conversion rate by that $Q$-factor in the Golden Rule.

The power from axion conversions incident on the cavity on resonance to a particular mode is given by \cite{Sikivie:1985yu}
\begin{align}
\label{eq:pag}
P_{a\rightarrow\gamma}~\approx~&10^{-22}\mathrm{W}
\left(\frac{C_{a\gamma\gamma}}{0.97}\right)^{2}
\times\\
&\left(\frac{V}{136~ \mathrm{L}}\right)\left(\frac{B}{6.8~
\mathrm{T}}\right)^{2}\left(\frac{\mathrm{C}_{nlm}}{0.4}\right)
\left(\frac{\rho_{a}}{\mathrm{0.45~  GeV/cm^3}}\right)\left(\frac{\nu}{\mathrm{650~MHz}}
\right)\left(\frac{Q}{\mathrm{50000}}\right),\nonumber
\end{align}
where $V$ is the cavity volume, $B$ is the axial magnetic field within the cavity (for simplicity assumed constant),
$\mathrm{C}_{nlm}$ is the form factor of the $nlm$ cavity mode, $C_{a\gamma\gamma}$ is the dimensionless
coupling of axions to photons (eqs.~\eqref{eq:lagrangian},\eqref{eq:cagg}),  $\nu$ is
the cavity resonant frequency, and $Q$ is the loaded cavity quality factor.  The axion energy density is given by $\rho_{a}$, chosen to be at a reference value of $0.45\,$GeV$/$cm$^3$ based on early work on galactic dark matter density inferences~\cite{Gates_1995}. While the energy density in the lab itself is not directly measurable, precision astronomical data is continuing to improve: on kiloparsec scales, typical measurements constrain the density to be in the range of $0.4$--$0.6$\,GeV$/$cm$^3$, while large-scale galactic inference come in slightly lower at $0.3$--$0.5$\,GeV$/$cm$^3$~\cite{de_Salas_2021}.

A small antenna inserted into the cavity extracts some of this RF power from
the TM$_{010}$ mode and delivers that power to
the input of a an ul\-tra-low-noise amplifier. 
The output of the first-stage amplifier is applied to a more-or-less
conventional receiver chain (either analog heterodyne or digital), and the bandwidth about the cavity resonance is
then Four\-i\-er-a\-nal\-yzed into a frequency spectrum in the search for an axion signal above the noise.
Axions thermally virialized with other galactic components would have a frequency
bandwidth of order $\frac{1}{2} m_a v_0^2$
where $v_0^2$ is the square of the virial velocity dispersion of
dark matter near Earth. For $m_a$ in the $\mu$eV range ($\sim$ 1 GHz), and
$v_0 \sim 10^{-3} c$, the axion line width is $\sim$ 1 kHz.
A reasonable normal metal loaded cavity $Q$ around $10^5$ has a cavity resonance bandwidth
$\sim$ 10 kHz. These considerations set the Nyquist parameters for
Fourier analysis of the cavity power spectrum. 

This idea was quickly applied to relatively small cavity experiments
at BNL~\cite{DePanfilis:1987dk}\ and the University of Florida (UF) \cite{PhysRevD.42.1297},
which demonstrated the experimental method and established the path for the current `second-generation' cavity experiments.  It was only in the late 2010's,
almost 30 years after the PQ axion was postulated, that ADMX Gen 2 finally demonstrated sensitivity to the
highly compelling `DFSZ' dark matter axion in a promising ax\-i\-on-mass region.
The modern realization of this technique is the focus of the remainder of this section.

\begin{figure}[ht]
\centering

\includegraphics[trim={0 0 0 0}, clip, width=.85\textwidth]{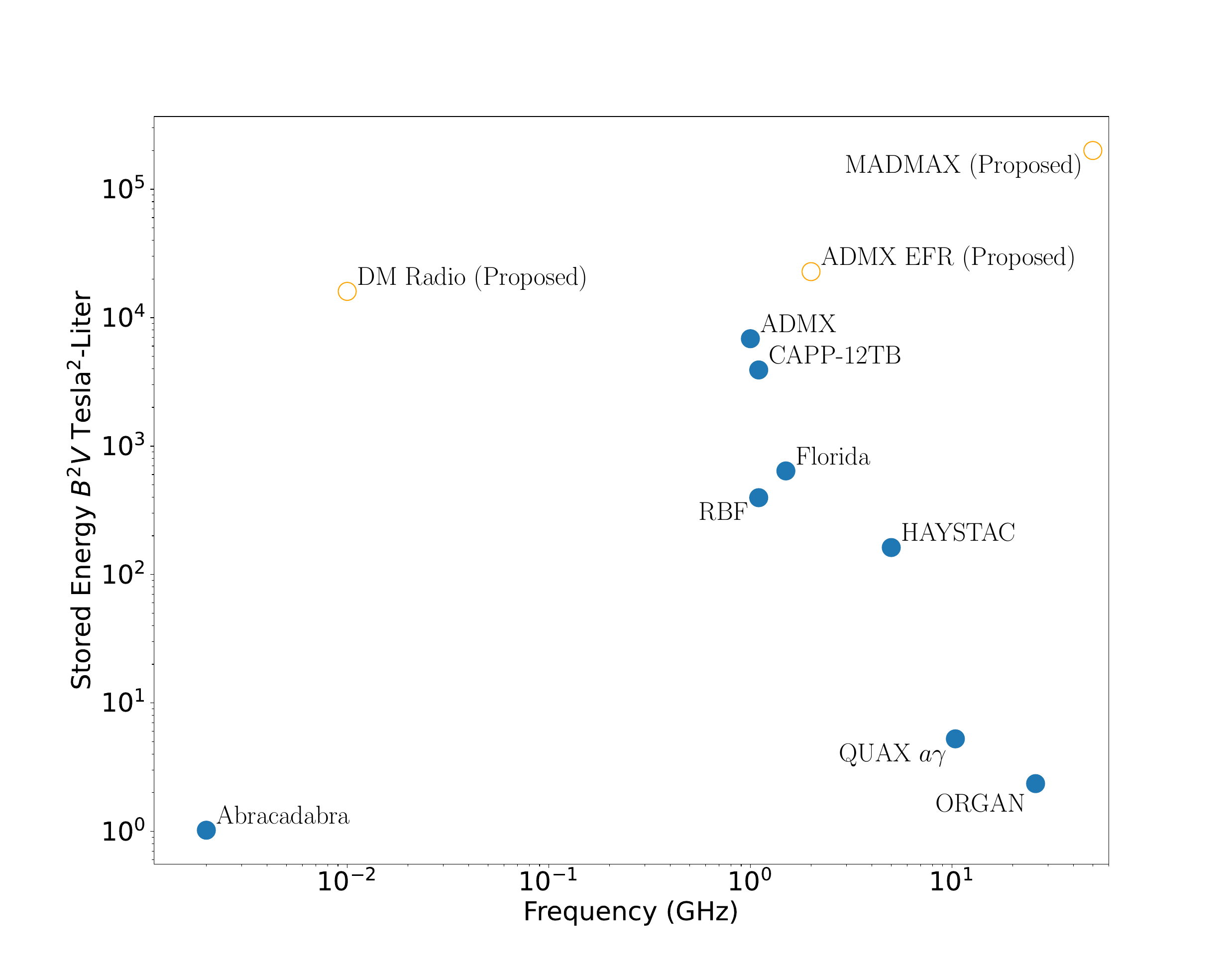}
\begin{tabular}{lSSl}
\hline
Experiment & {Volume (L)} & {Field (T)} & Reference \\
\hline
Florida           &    10  &  8   & \cite{PhysRevD.42.1297} \\ 
ADMX      &          119  &  7.6    & \cite{doi:10.1063/5.0037857}  \\ 
CAPP-12TB  &          37 &   10.3 & \cite{https://doi.org/10.48550/arxiv.2210.10961} \\ 
HAYSTAC       &         2   &  9   &  \cite{PhysRevD.97.092001} \\ 
ORGAN         &      0.048   & 11.5  & \cite{doi:10.1126/sciadv.abq3765} \\ 
QUAX $a\gamma$  &     0.08  &  8.1  & \cite{PhysRevD.103.102004} \\ 
Abracadabra  &         1.02   & 1.0  &  \cite{PhysRevLett.122.121802}  \\
RBF              &  11  &   6   &  \cite{PhysRevLett.59.839} \\
ADMX EFR (Proposed)&  258 &      9.4  & \cite{Adams:2022pbo} \\ 
DM Radio (Proposed)& 1000 &   4   & \cite{PhysRevD.106.103008} \\
MADMAX (Proposed)   &    2000 &  10 &   \cite{MadmaxWhitepaper} \\ 
\end{tabular}
\caption{The total magnetic stored energy ($B^2 V$) for a selection of haloscopes discussed here,
spaced as a function of frequency.
Solid blue circles are experiments with published results and the volume is for the resonator used in the reference.
Hollow orange circles are magnets described in white papers of proposed experiments,
and use the total magnet bore volume, not the volume of a particular resonator.
The speed at which frequencies can be scanned is proportional to the stored energy squared.
Note that the trend so far has been to use smaller bore,
higher field magnets with lower stored energies at higher frequencies.}
\label{fig:bsquaredv}
\end{figure}

The number and variety of haloscopes, operating,
proposed and conceptualized, continues to increase,
so this review will necessarily be incomplete
by the time of publication. We discuss in detail the development of the oldest operating experiment, ADMX, and provide a limited selection of other notable efforts.

\subsubsection*{ADMX Overview}

The Axion Dark Matter eXperiment (ADMX) is
the long\-est-run\-ning haloscope and by far has scanned the
widest mass-range at DFSZ sensitivity.
ADMX was the first experiment to achieve sensitivity to even weak\-er-than-DFSZ couplings in the
few $\mu$eV ax\-i\-on-mass range.
ADMX began operation in the 1990's, initially sited at Lawrence Livermore National Laboratory,
and built as a conceptual fol\-low-on to the
UF pilot haloscope \cite{PhysRevD.42.1297}.
It was recognized that the early BNL and UF pilot experiments
lacked the
sensitivity required to detect plausible QCD dark-mat\-ter axion models by several or\-ders-of-mag\-ni\-tude.
The initial ADMX was a scale-up of the UF design, but
with a significantly greater cavity volume.
The first generation of ADMX (``Phase 0 ADMX'') used
cryogenic transistor first-stage amplifiers (``HFETs'').
Although considered low-noise in the general microwave community,
these cryogenic HFET amplifiers nonetheless have electronic noise considerably
greater (perhaps by a factor of 100) than that allowed by the standard quantum limit (SQL)
of $h \nu/2$, with $\nu$ the signal frequency.
Nonetheless, because of the large cavity volume,
this ``Phase 0'' ADMX achieved sensitivity to the benchmark KSVZ axion
for masses in the plausible range of the QCD dark-mat\-ter
axion \cite{PhysRevLett.80.2043}.
Unfortunately, the DFSZ coupling is almost a factor of 10 weaker in developed cavity
signal than that of the KSVZ coupling, and the ADMX Collaboration and others
certainly appreciated that what the community called a
``definitive search''\footnote{The term ``definitive search'' is
somewhat hyperbolic in that a motivated theorist could
certainly conjure a scheme for axions to evade it. Furthermore, the sensitivity to the coupling depends on the assumption about the local dark matter density on Earth, which is not directly determined and is order-one uncertain in standard cosmological models.}\
requires sensitivity to DFSZ axions.

Recall, of the parameters that contribute to the haloscope sensitivity [eq.~\eqref{eq:pag}],
the magnetic field is perhaps the most obvious parameter one could improve.
However, the ADMX Phase 0 magnet is of NbTi conductor construction, and
its central field around 8 T is near the
maximum practical field for NbTi technology at 4.2 K.

Returning to the parameters contributing to the sensitivity,
another promising factor to improve is the system noise (see also section~\ref{sec:futurecavity}).
In principle, achieving the Standard Quantum Limit (SQL) of noise
would improve the sensitivity by a factor of $\times$100 relative to ADMX Phase 0 
at fixed scanning speed and other parameters,
though in practice some of the improvement in amplifier noise
is lost in downstream electronics and some of the the
sensitivity improvement is applied instead on speeding up the search.
Nonetheless, it was appreciated during the epoch of
ADMX Phase 0 that approaching the SQL in noise would
allow ADMX to achieve DFSZ sensitivity over a wide ax\-i\-on-mass range.

Although the idea of there being a lower quantum limit to noise
of an amplifier is credited to Heitler in the 1930's~\cite{heitler1984quantum},
it was not until the development of the maser that such
low mi\-cro\-wave-amp\-li\-fi\-er noise could begin to be explored in the laboratory.
Throughout the 1960's, low-noise maser microwave amplifiers were constructed,
giving hints of measurable quantum noise buried in the am\-pli\-fi\-er-noise output.
However, the maser amplifiers themselves were not typically operating
close to the SQL, they were considerably noisier.
Although maser amplifiers were deployed in ra\-dio-tele\-scope applications,
ultimately, for these frequencies, many ra\-dio-tele\-scope receivers transitioned to
cryogenic HFET transistor amplifiers by the 1990's.
About that time, amplifiers based on dc Superconducting
QUantum Interference Devices (SQUIDs) were demonstrating
impressively low noise up to perhaps low-UHF frequencies.
However, the standard ``Ketchen'' SQUID fabrication geometry
has a parasitic capacitive coupling between the SQUID
input loop and the SQUID washer, and this capacitance rolls-off the
SQUID gain with increasing frequency.
Early in the 2000's, it was realized that a modification
of this geometry, in which the input loop
is replaced by a resonant stripline coupled to the SQUID loop, would allow the parasitic capacitance
to be cancelled by the reactance of the input stripline operating
slightly off resonance \cite{10.1063/1.124486}.
This concept is distantly related to an old ra\-dio-elec\-tron\-ics
idea of operating a transmitter slightly off of the an\-ten\-na-feed resonant
frequency to cancel parasitic impedances, the so-called
``shunt detuned position''.
Such Microstrip SQUID Amplifiers (MSAs) demonstrated low noise, nearly
down to the SQL at
signal frequencies up to at least several GHz. This advance
provided the enabling technology for ADMX to achieve DFSZ sensitivity.

Incorporating MSAs into ADMX introduced several technical challenges.
First, MSAs, being based on dc SQUID technology, are disturbed by
even very tiny ambient magnetic fields. For the MSA to operate near the
ADMX RF cavity, it must be shielded from the 8 T field down to field strengths much less than a gauss.
Since the mag\-net\-ic-fi\-eld gradient near the MSA is high, the resulting force
on a mag\-net\-ic-shi\-eld system would be enormous, perhaps several tons.
Restraining such forces in a deep cryogenic environment is a challenge.
To deal with these forces, the fi\-eld-can\-cel\-la\-tion coil surrounding the SQUID is co-ax\-i\-al\-ly mounted on a mandrel with a
coun\-ter-wound coil energized in in opposite phase. The two coils are designed
to have net-zero force.\footnote{Net-zero force implies net-zero mutual inductance
between the coil pair and the main magnet. This is highly desirable: should
one of the coils quench the other would not have large induced currents.}
Second, achieving low MSA noise requires operation  at di\-lu\-tion-re\-frig\-er\-a\-tor temperatures.

While cryogenic HFET transistor amplifiers no longer much improve in noise
below physical temperatures of a few K, the MSAs continue to improve in noise all the way down to
ambient temperatures approaching $h \nu$ (around 50 mK at 1 GHz).
Hence, in practice an ADMX incorporating MSAs requires surrounding the cavity and
MSA electronics in ``1 K'' thermal shield outside of the di\-lu\-tion-re\-frig\-er\-a\-tor
space, which called for a significant re-de\-sign of the ADMX experiment insert.
A version of ADMX incorporating the fi\-eld-free region for the MSA plus
di\-lu\-tion-re\-frig\-er\-a\-tor was proposed. A review deemed the engineering required for operating the MSA in the fi\-eld-free
region to be high risk, and therefore the project would proceed in two sequential
stages: ``Phase 1a'' ADMX would demonstrate the MSA could operate successfully near
the 8 T main magnet in its fi\-eld-free region, and a fol\-low-up ``Phase 1b'' would
retrofit the dilution refrigerator.

The Phase 1a ADMX successfully demonstrated MSA operation at pumped
$^4$He physical temperatures
around 1 K \cite{PhysRevLett.104.041301}.
The resulting system noise
was somewhat improved over that of the HFET ADMX Phase 0. With
MSA operation demonstrated, ADMX proposed retrofitting the dilution
refrigerator, thereby lowering the cavity and amplifier physical
temperature to around 100 mK and realizing the low-noise potential
of the MSA in a production search, approved in the DOE-HEP  ``Generation 2''
dark-mat\-ter-de\-tec\-tion call. The di\-lu\-tion-re\-frig\-er\-a\-tor ADMX Phase 1b was
renamed ``ADMX Gen 2''. The ADMX Gen 2 project moved to the University
of Washington into a newly refurbished facility. After several preparatory
data runs without the dilution refrigerator, the dilution refrigerator
was successfully deployed in what was internally called within ADMX
``run 1A''. Run 1A was notable for achieving sensitivity to DFSZ
axions over a plausible range of QCD dark-matter axion masses,
a long-time goal of axion research
finally realized \cite{PhysRevLett.120.151301}.
The fol\-low-up run 1B significantly
extended the mass range and demonstrated long-term, stable operations
at better than DFSZ sensitivity \cite{PhysRevD.103.032002}.
The current Run 1C continues scanning and
in the near-term run 1D will finish the single-ca\-vi\-ty ADMX operations.
The fol\-low-up scan is run 2A, which will deploy multiple RF cavities
to allow a searching at higher axion masses while maintaining good sensitivity.

The various phases of ADMX are described in the
literature \cite{sciamadmx}.
The section below expands on details of the current ADMX Gen 2.

\subsubsection*{ADMX ``Gen 2''}

\begin{figure}[ht]
\centering
\includegraphics[width=0.5\textwidth]{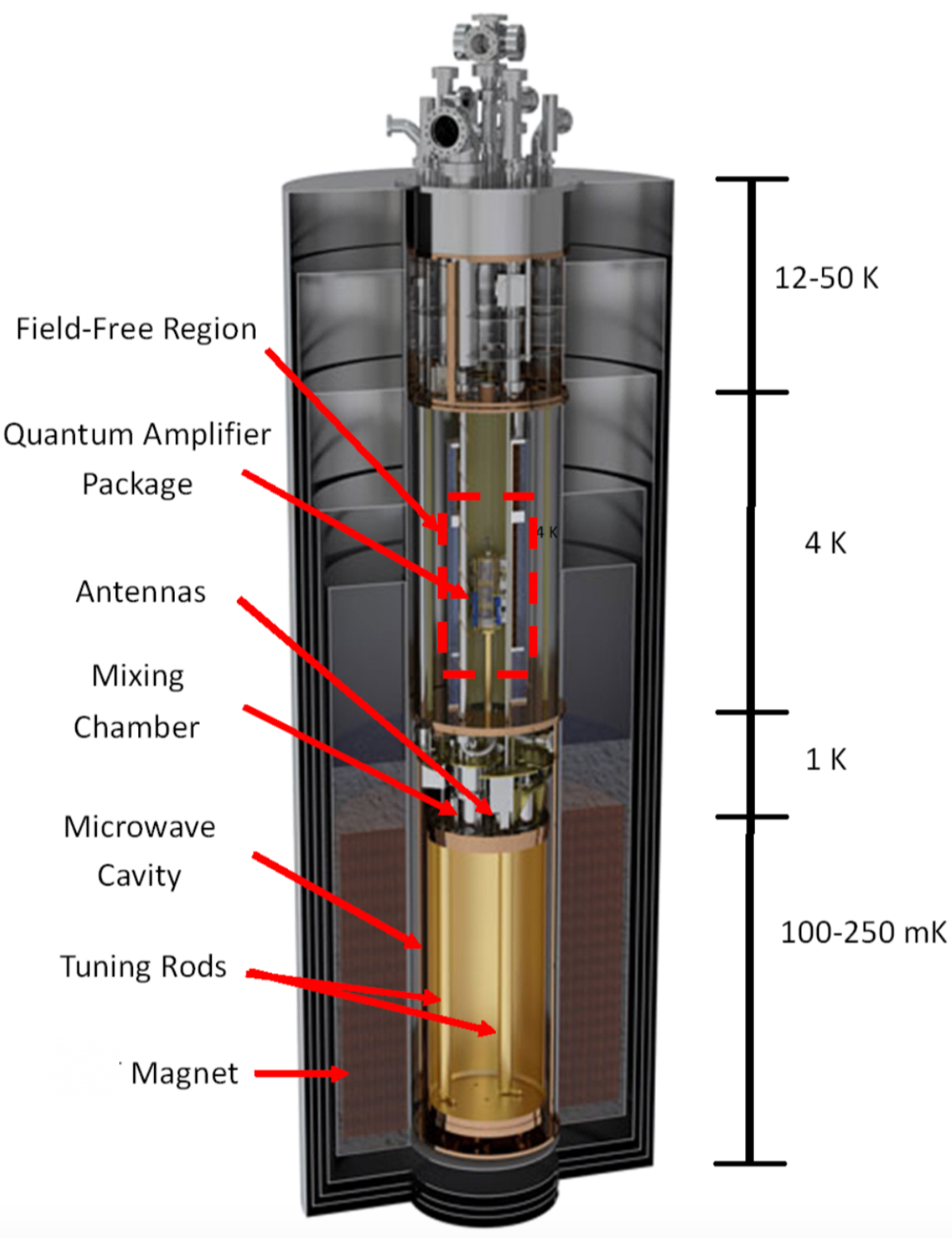}
\caption{Sketch of the ``Gen 2'' ADMX experiment insert. The RF cavity near the bottom
is in the bore of a large superconducting solenoid magnet. Above the cavity
are di\-lu\-tion-re\-frig\-er\-a\-tor and quan\-tum-am\-pli\-fi\-er components.
The quan\-tum-am\-pli\-fi\-er
components are situated within a mag\-netic-fi\-eld-free region established with a ``bucking coil.''
From \citen{doi:10.1063/5.0037857}.}
\label{fig:ADMX-experiment-diagram}
\end{figure}

The cur\-rent\-ly-oper\-a\-ting phase of ADMX is ADMX Gen 2
(``ADMX'' in what
follows)\cite{doi:10.1063/5.0037857}.
ADMX is located at the Center for Experimental Physics and Astrophysics (CENPA)
at the University of Washington, Seattle.
The main experimental package, the ``insert'' is located in the bore of
a 8 T superconducting solenoidal magnet, see figure~\ref{fig:ADMX-experiment-diagram}.
The insert consists of a right-cir\-cu\-lar cylindrical microwave cavity
(OFHC cop\-per-plated over stainless steel, approximately 1/2 m diameter, 1 m tall),
in which are two movable tuning rods and an ad\-jus\-ta\-ble-depth antenna.
Above the cavity are components of a dilution refrigerator (for cooling
the RF cavity and cryogenic microwave electronics), mo\-tion-con\-trol
components (for rods and antennae).
Yet higher up is a li\-quid-he\-li\-um reservoir and the cryogenic microwave
electronics.
In addition to supplying cooling to the insert, the reservoir contains a
smaller superconducting solenoid magnet (the ``bucking coil'') to ``buck'' (zero) the magnetic field
in the vicinity of the cryogenic electronics.
Above the reservoir is the cold head of a pulse-tube cooler that establishes
a 50 K thermal shield.
In addition, there are two pumped ${^4}$He refrigerators (``pots''), one for
the 1.5 K thermal shield surrounding the cavity and electronics,
the other condenses the circulating ${^3}$He-${^4}$He mixture of the dilution refrigerator.
A substantial portion of the ADMX infrastructure also includes a dedicated
Linde L1410 helium liquefaction system for liquefying helium exhaust gas from the main-mag\-net and re\-ser\-voir.

Recall, the enabling technology for ADMX is the quantum amplifier front end of the
cryogenic electronics package.
The most recent incarnation of ADMX uses a Josephson Parametric Amplifier (JPA)
in a ``cur\-rent-pump'' mode, fabricated in alu\-mi\-num,
which is based on dc SQUID technology.
These devices provide around 20 dB of microwave power amplification with
near quan\-tum-li\-mi\-ted
noise.
JPA's are tunable and dissipate very little power from the bias current. However, since they
contain Josephson junctions, they need be shielded from the 8 T main field
by the ``bucking coil''.
The output of the JPA feeds a high-elec\-tron mobility cryogenic JFET amplifier (``HFET amplifier''), at
which point there is enough low-noise gain for the signal to exit the cryostat.
The low-noise gain offered by the JPA is critical: recall that for a fixed scanning time, the
sensitivity of the experiment degrades linearly with the JPA noise, and for fixed
sensitivity, the scanning time increases quadratically with JPA noise.
Without these quantum amplifiers, a search at DFSZ sensitivity over the
plausible QCD dark-mat\-ter ax\-i\-on-mass range could take centuries.

\begin{figure}[t]
\centering
\includegraphics[width=0.8\textwidth]{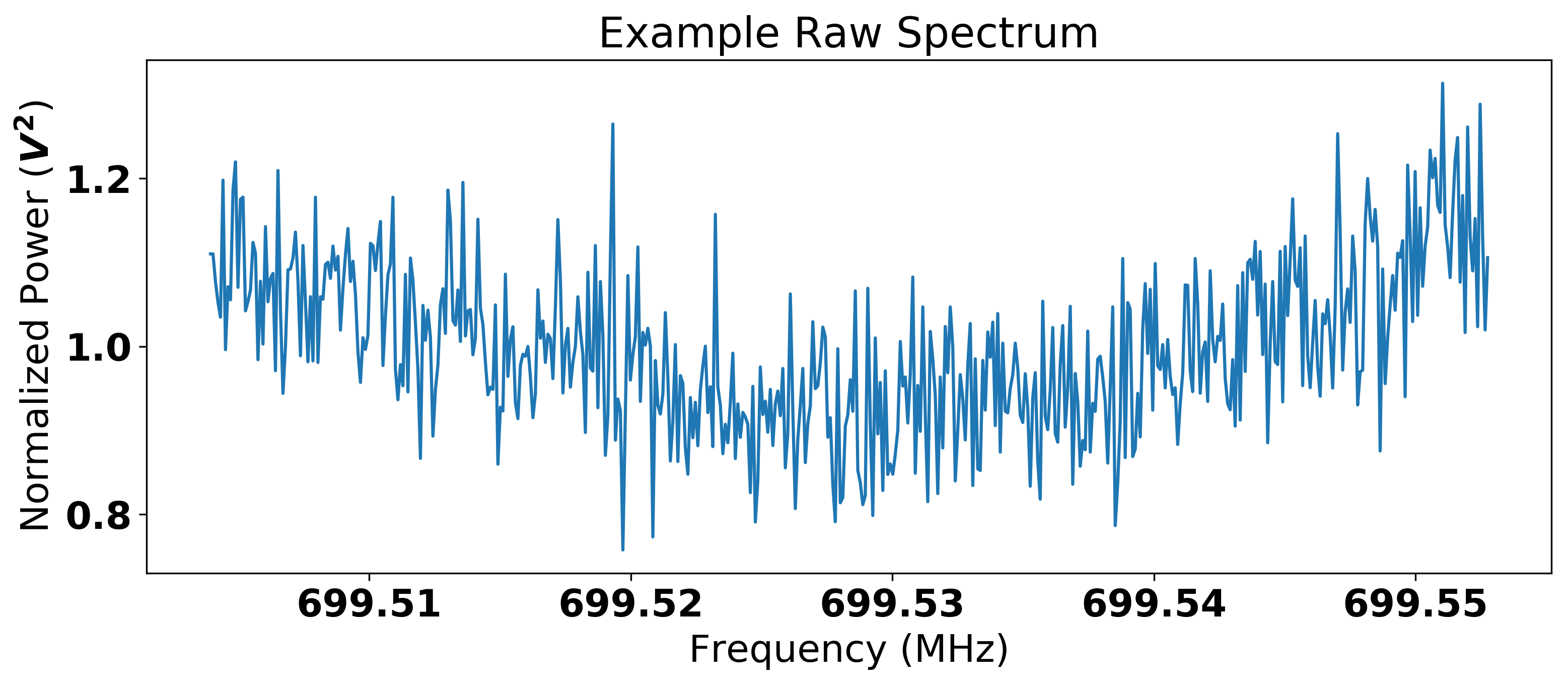}
\caption{A single ``raw'' unprocessed power spectrum in ADMX.
The overall shape is due to the interaction of the cavity, transmission lines,
and device inputs.
From \citen{PhysRevD.103.032002}.}
\label{fig:ADMX-raw-spectrum}
\end{figure}

The microwave signal as well as slow-con\-trols signals are processed by
room-tem\-per\-a\-ture data-acqui\-si\-tion electronics.
Here, the microwave signal is mixed-down to lower IF frequencies and digitized.
The resulting vol\-tage-time series is processed into a
power spectrum in the vicinity of the cavity resonance.
One power spectrum results from approximately each minute of acquired time series.
The time-ser\-ies is over-sam\-pled, so spectral resolution can be adjusted af\-ter-the-fact to,  e.g.,
match the isothermal axion phase space.
The achievable frequency resolution is ultimately limited by the stability and noise of
the local oscillators and digitizers, around fractional frequency resolution of $10^{-14}$.
A typical single power spectrum is shown in figure~\ref{fig:ADMX-raw-spectrum}.
The overall shape is due to the interaction of the cavity, transmission lines,
and device inputs.
This overall shape is understood and the analysis searches for relatively narrow small-am\-pli\-tude peaks,
with fractional width around $10^{-6}$ (for a virialized axion), on top of
the overall structure.  The experimental cadence is to record such a spectrum on the timescale of minutes, then step the cavity frequency
by a small amount (relative to the cavity resonance width), then record another spectrum,
then step the cavity frequency, etc.
Months of such data are collected and used in a search for axion signals.

\begin{figure}[ht]
\centering
\includegraphics[width=0.8\textwidth]{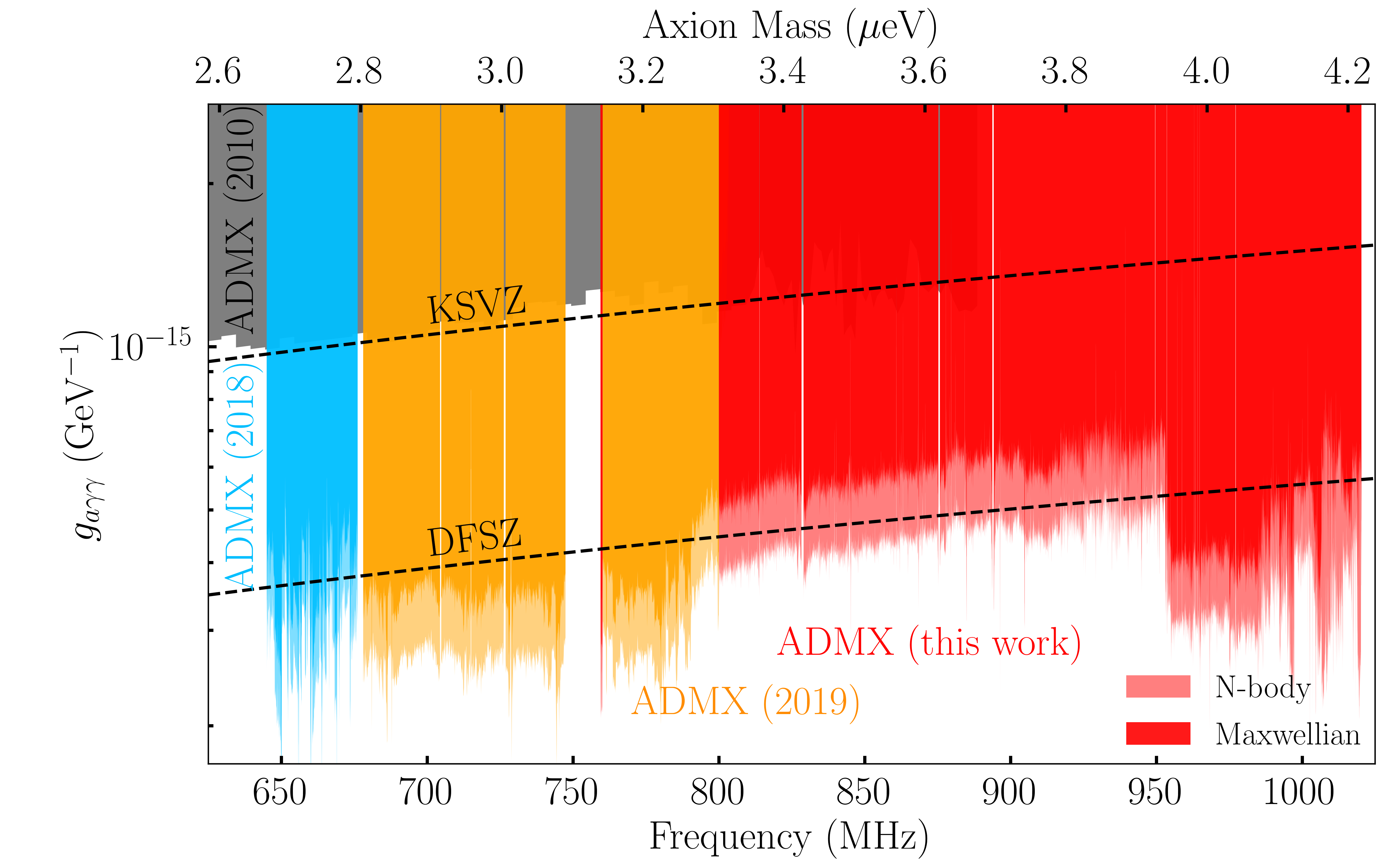}
\caption{
A recent set of ADMX limits.
The horizontal axis is axion mass (upper) or frequency (lower),
and the vertical axis is the coupling of the axion to two photons.
The two diagonal lines indicate benchmark KSVZ and DFSZ axion models.
The grey region shows very early ADMX limits at KSVZ sensitivity.
The blue shows the later ADMX limit;
it is notable for achieving
DFSZ sensitivity, a long-term goal is axion research.
The more recent orange and red limits are also shown.
The dark limits assume an isothermal sphere axion phase-space and $\rho_{\mathrm{DM}}=0.45\, \mathrm{GeV}/{\mathrm{cm}^3}$.
The lightened regions assume a DM density and phase space distribution extracted from N-body simulations
of  Milky-Way-type galaxy structure formation~\cite{lentznewsignal}.
From \citen{PhysRevLett.127.261803}.
}
\label{fig:ADMX-sensitivity}
\end{figure}

A recent set of ADMX limits is shown in figure~\ref{fig:ADMX-sensitivity}, where the
horizontal axis is axion mass (upper) or frequency (lower),
and the vertical axis is the coupling of the axion to two
photons \cite{doi:10.1063/5.0037857}.
The two diagonal lines are the benchmark KSVZ and DFSZ axion models.
The blue region shows very early ``Phase 0'' ADMX limits at KSVZ sensitivity,
representing almost a decade of da\-ta-tak\-ing.
The orange shows the later ADMX ``Run 1A'' limit, representing
about a year of da\-ta-tak\-ing. The orange region is notable for demonstrating
DFSZ sensitivity, a long-term goal of axion research.
The yet more recent red ``Run 1B'' limits, also at better than
DFSZ, represents about year of da\-ta-tak\-ing and shows a considerable
speed-up. ADMX is near finishing taking data in ``Run 1C'' and is
transitioning to ``Run 1D''.
One subtlety: on close examination, note there are two limits, e.g., light
and dark red.
The darker color limit assumes an isothermal sphere axion phase-space,
the light color assumes the phase space from the N-body model
of structure
formation~\cite{lentznewsignal}
and is likely the more realistic. 

ADMX continues to take data. The ADMX strategy is to search upwards in frequency.
A major upgrade will occur when ADMX is retrofitted with a 4-ca\-vi\-ty insert package (``Run 2A'').
The smal\-ler-di\-a\-me\-ter of each cavity moves the search to
higher frequencies.
Approximately a year after that, ADMX will be fitted with a yet high\-er-fre\-quen\-cy
package (``ADMX-EFR'') to allow the search to move into yet higher axion
masses~\cite{Adams:2022pbo}.

\subsubsection*{CAPP/CULTASK}

There is an ambitious program that includes RF-cav\-i\-ty axion research underway at the
Center for Axion and Precision Physics Research (CAPP) in
Korea \cite{PhysRevLett.126.191802}.
This program has several platforms for axion searches and R\&D.
The flagship platform is CAPP's Ultra-Low Temperature Search in Korea (CULTASK).
The conversion magnet is a 12 T peak-fi\-eld solenoid.
Inside the magnet bore is an approximately 30 liter conversion cavity (radius 0.132 m, height 0.560 m)
with a single metallic tuning rod (radius 0.034 m) tuned with piezoelectric actuators.
The cavity and cryogenic electronics are cooled with a di\-lu\-tion-re\-fridg\-er\-a\-tor system
operating at temperatures below 50 mK.
The front-end of the cryogenic RF electronics is a Josephson parametric amplifier.

\begin{figure}[t]
\centering
\includegraphics[width=0.85\textwidth]{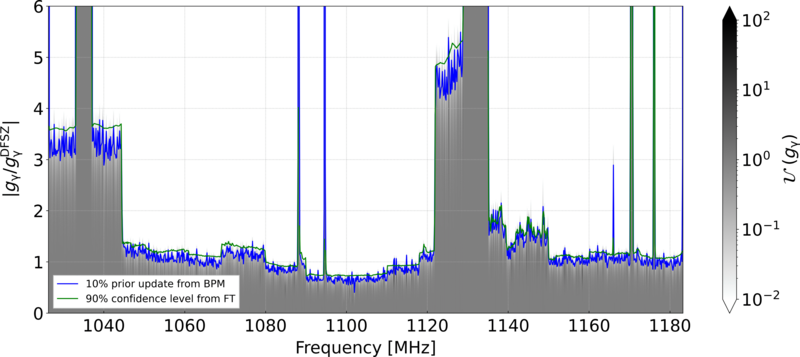}
\caption{
Recently excluded CAPP axion parameter space.
From \citen{PhysRevX.14.031023}.
}
\label{fig:cultask}
\end{figure}

Importantly, this system has reported a result at DFSZ sensitivity over a relatively narrow
mass range \cite{PhysRevLett.130.071002}.
The first 20 MHz wide da\-ta-set was taken
above 1 GHz (approximately 4.5 $\mu$eV axion mass) in March 2022. The CULTASK system continues to take data and has recently greatly expanded its DFSZ-sensitive region \cite{PhysRevX.14.031023}.

The broader CAPP program also reported results from low\-er-mag\-net\-ic-fi\-eld platforms at
axion couplings above DFSZ at a variety of masses. The program includes R\&D on superconducting cavities
and ``mul\-ti-cell'' (mul\-ti-ca\-vi\-ty)
systems \cite {PhysRevLett.125.221302}.

\subsubsection*{HAYSTAC:}~
The Haloscope at Yale Sensitive to CDM (HAYSTAC) experiment is a relatively
small-vol\-ume haloscope with a  25.4 cm long and 10.2 cm inner diameter main cavity with a  5.1 cm diameter tuning rod, in an 8 T magnetic field~\cite{PhysRevD.97.092001,Backes2021}.
The experiment first took data with with a single Josephson parametric amplifier and reported an exclusion result for masses in the range of 23.15--24.0$\mu$eV and couplings a factor of 2.7 above the KSVZ line~\cite{Brubaker:2016ktl,PhysRevD.97.092001}. An upgrade to Phase II then covered the mass range 16.96--17.12 and 17.14--17.28 $\mu$eV in Phase IIa~\cite{Backes2021} and 18.44--18.71 $\mu$eV in Phase IIb for axion-photon couplings a factor of 2 above KSVZ~\cite{HAYSTAC:2023cam}.
Of particular interest in this second phase is that the receiver used vacuum
squeezing to evade the standard quantum noise limit \cite{Backes2021}.
In this case, the squeezing improved the SNR by around $\sqrt{2}$,
thereby speeding up the scan by around a factor of 2.
This is the first demonstration that state-squeez\-ing could be
used to improve haloscope sensitivity (see section~\ref{sec:futurecavity}).

\subsubsection*{RADES:}

The Relic Axion Dark-Matter Exploratory Setup (RADES) is a haloscope axion search
within the bore of the CERN Axion Solar Telescope (CAST)
helioscope\cite{alvarezmeconrades}.
The cavity package consists of five right-cir\-cu\-lar cylindrical cavities
in series, coupled at each end with a\-per\-ture-fil\-ters.
The cavities were cooled to around 1.8 K and read out by a cryogenic
HEMT amplifier. The magnetic field in the neighborhood of the cavity
was around 8.8 T. A result was reported for this
system at an axion coupling two orders stronger than the CAST bound \cite{alvarezmeconrades}.
What is interesting about this result is the frequency is around 8.4 GHz (34 $\mu$eV)
and it is now one of the most sensitive direct searches at these relatively high axion masses.

\subsubsection*{ORGAN:}

The Oscillating Resonant Group AxionN (ORGAN) is a haloscope aimed at high frequencies
(high axion masses) in the range 15-50 GHz (62-207 $\mu$eV).
The magnet had peak field around 11.5 T and a relatively small cavity volume.
ORGAN reported an early result for axion masses in the
range 63-67 $\mu$eV \cite{doi:10.1126/sciadv.abq3765}.
Although the sensitivity is for axion coupling far from DFSZ, this
is a demonstration of the haloscope technique to what had previously been
inaccessible axion masses.

\subsubsection*{QUAX:}~
The (QUAX) collaboration ~\cite{Barbieri:2016vwg} was conceived to search for the axion-electron coupling with a resonant experiment.  Here, a small RF cavity is loaded
with a ferromagnetic material. Due to the relative motion of the Earth and axion dark-mat\-ter
halo, there is an axion gradient interaction which leads to an effective mag\-net\-ic-fi\-eld
elec\-tron-spin
interaction. The electrons precess and this, in effect, transfers power from the axion field
to the precessing electrons. This resonant power is then re-emit\-ted as electromagnetic radiation inside
a tuned RF 
cavity \cite{Crescini2018}.
A recent QUAX apparatus consisted of a small, non-cy\-lin\-dri\-cal cavity, loaded with GaYIG spheres,
with relevant mode frequency near 14 GHz.  The system was cooled to mK temperatures and read out with a JPA.
The resulting limits were in a narrow axion mass band near 43 $\mu$eV with
coupling sensitivity still far from the QCD axion coupling. \cite{PhysRevLett.124.171801}

In the course of developing resonators, the QUAX group has also run searches for the axion -photon coupling as a standard haloscope.  Recently, they have demonstrated near-KSVZ sensitivity to dark matter axions in a narrow band around 34 $\mu$eV in a system held at 20mK in a 7 T field.\cite{PhysRevD.110.022008}

\subsection{Next Steps for Cavity Haloscopes}
\label{sec:futurecavity}

The prospects for improving axion haloscopes with new technologies are good, and can be divided into: larger magnets, lower noise, and improved resonators.

The rate that frequency bandwidth can be scanned is proportional to
\begin{equation}
\frac{d\nu}{dt} \propto Q B^4 V^2 \frac{1}{{T_s}^2},
\end{equation}
 where $T_s$ is the system noise temperature (substantially the sum of the amplifier noise
and the cavity physical temperature) and $Q$ is the linewidth of the cavity resonance.
In particular, notice the importance of the experimental
parameters $B, \,V,$ and $T_s$ in the scan rate.
For a practical search, the physical temperature of the cavity must be very low,
the first-stage amplifier must therefore be of extremely low-noise design,
and the magnet field strength and volume should be as large as possible.  Note that the cavity bandwidth increases as the $Q$ decreases so improvements in cavity $Q$ offer less dramatic improvements in scan time (or sensitivity) compared to increased $B$ or reduced $T_s$.  The $Q$ dependence also becomes more complicated when the cavity linewidth is narrower than the axion linewidth from the velocity dispersion of dark matter: tuning schemes can still gain scan speed, but at the expense of becoming insensitive to axion linewidths narrow er than that of the cavity.

\subsubsection*{Magnets}

    The power as measured by an axion haloscope is proportional to the active volume of the resonator, and to the average magnetic field squared.  This translates directly to the stored energy of the magnet: Bigger is better.  The stored energy of axion haloscope magnets has been steadily increasing over time, as has their cost.  The highest field used to date is 18T with a 70mm bore, 500 mm long~\cite{PhysRevLett.128.241805,PhysRevD.106.092007}.
    
    The stored energy for several of the experiments discussed here is shown in figure~\ref{fig:bsquaredv} for comparison.  With existing technology, there is a decision for future axion haloscope magnets:  an experimental group could choose to use Nb$_3$Sn superconducting technology to build a magnet with a very high field over a very small volume, or the group could choose a lower field magnet with NbTi superconductor but a much larger volume. 
    
    One could conceive of large-bore magnets
wound with
Nb$_3$Sn superconductor, a considerably more expensive, fragile and therefore riskier superconductor;
the resulting magnet would be perhaps a factor
of 10 more expensive per unit of stored energy than that of NbTi.
Hence, large volume 20 T magnets are immediately achievable, but at very high cost, and are being investigated \cite{Gupta2019}. 
Yet-high\-er magnetic fields are
conceivable with high-T$_c$ superconductors, perhaps even magnets into the 40 T
range \cite{Hahn2019}. There is considerable interest in this high-T$_c$ technology,
but the engineering challenges have not yet been surmounted
for building high-T$_c$ solenoid magnets with large-vol\-ume bores and
robust quench-pro\-tec\-tion. One could perhaps envision, though, such magnets
eventually being in\-ex\-pen\-sive-enough and technically ma\-ture-enough to be
deployed in a Si\-ki\-vie-ca\-vi\-ty search; this would be a potential
``game changer'' in haloscope design.

\subsubsection*{Noise:}
    
    The current generation of axion haloscopes operates near the standard quantum limit with a noise temperature equivalent $kT\simeq h\nu/2$ \cite{PhysRevD.26.1817}.  This limit arises from the use of linear amplifiers, which must add a half-photon of noise in amplification to preserve the Heitler phase/pho\-ton-num\-ber (power) uncertainty relation.  Two directions are being pursued to avoid this bound, both based around trading larger phase uncertainty for lower power uncertainty.  
    
    The first of these is ``squeezing'', in which a special noise source that has ``squeezed'' most of the noise into one of the ``quadratures'' provides the background, and the system is read out and amplified in the other quadrature.  Interestingly, this cannot directly lower the noise of an axion haloscope; the fundamental noise from the cavity physical temperature cannot be squeezed.  What can be done, however, is to over-couple to the detector cavity.  In this case, both the signal and the noise from the cavity are reduced, but the cavity covers a larger range of frequencies.  The over-coupled antenna also has a noise component that is reflected off the cavity; this is where the squeezed source is used.  Thus, in principle, a wider range of frequencies can be explored at once with squeezed noise than with a white noise source, and thus frequencies can be scanned more quickly~\cite{PhysRevD.88.035020,zheng2016accelerating}.  As mentioned above, this has already been successfully demonstrated in the HAYSTAC experiment \cite{Backes2021}, with a factor of 2 improvement in scan speed.  The limiting factor in this scheme is that any noise from losses between the cavity and amplifier are magnified by the squeezing, and the cavity and amplifier are necessarily separated, both because of the need for intervening RF components and because the amplifier needs a very low-field in which to operate.  Improvements in low-loss and fi\-eld-tolerant RF components amplifiers are needed for further improvement.

    The second approach to a post-SQL axion search is to use a non-demolition measurement where the number of photons in the cavity is measured, and the phase information is not.  This is well-aligned with present efforts to develop quantum computing hardware and has significant technological overlap.  The primary challenge is to couple photons from the detection cavity in a high field, to a readout cavity in zero field to accommodate the readout electronics.  A single mode transport system has been demonstrated in a fi\-eld-free system~\cite{Dixit:2020ymh}.  A more complicated system that may avoid some of the transport issues has been proposed in \cite{PRXQuantum.2.040350}, and a successful demonstration could potentially lead to an orders of magnitude increase in scan speed.
    
\subsubsection*{Resonators:}

    The principal difficulty in the half-wave\-length resonators used presently for axion haloscopes is that at higher frequencies, the small size of the resonators and reduced quality factor makes any dark matter signal too feeble to be useful.  A number of solutions are in development, as noted below.
    
\subsubsection*{Multicavity Systems:}
    
    For nonrelativistic axion dark matter, the Compton wavelength is much smaller than the de Broglie wavelength.  The signals produced from multiple haloscope cavities within the axion de Broglie wavelength will be in-phase, so this suggests that one path to a more sensitive haloscope is to combine the signals from multiple cavities.  This is the proposed technique to utilize a magnet with characteristic dimensions larger than that for a single cavity in the ADMX G2 and ADMX EFR experiments \cite{Adams:2022pbo}.  There have also been efforts made to combine the field from multiple cavities in more complicated geometries to achieve the same effect.  Both cases present a challenge that all cavities must be tuned relatively closely in frequency to coherently combine the signals.
    
\subsubsection*{High-Q Cavities:}

    For resonant haloscopes, the rate at which frequencies can be explored at constant sensitivity is proportional to the resonant quality factor $Q$ \cite{PhysRevD.103.032002}, which is limited by the loss of the cavity walls.  At present, high-purity copper cavities with unloaded $Q$s as high as
    $\sim$100,000 at 1 GHz have been used in haloscopes~\cite{doi:10.1063/5.0037857}, and $Q$ scales as $\nu^{2/3}$ at higher frequencies.  Much higher quality factors (as high and more as $10^{11}$) have been achieved in superconducting cavities, but the superconductor tends to become quite lossy when immersed in magnetic fields of the strengths used in axion experiments.  Nevertheless, there have been some reports of cavity geometries that could maintain $Q$'s a factor of 6 better than copper even in multi-Tesla fields using Nb$_3$Sn \cite{Posen:2022tbs} and high-Tc superconductors~\cite{Ahn:2021fgb, Ahn:2019nfy}.  If these methods can be successfully implemented in a full-scale axion search, it would enable higher frequencies to be explored on tractable timescales.

\subsubsection*{Multiwavelength Resonators:}
    
    In contrast to coherently combining the signal from multiple resonators, one can envision a single resonance that spans multiple wavelengths.  For naive geometries, multiwavelength resonances overlap very poorly with the axion wave.  However there have been efforts to use dielectrics to shape the electromagnetic resonance to have significant overlap with the axion wavefunction for microwave frequencies above 10 GHz using periodic dielectric structures such as in the ORPHEUS experiment \cite{Cervantes:2022yzp,Cervantes:2022epl}, and in the far infrared using thin film deposition with the LAMPOST experiment in the context of dark photon searches~\cite{PhysRevD.98.035006, Chiles:2021gxk}.  It has also been suggested to use subwavelength metallic structures to emulate the wavelength of photons in a plasma to produce larger volume resonators for a given frequency (the Plasma Haloscope \cite{Lawson2019, ALPHA:2022rxj}).

    Of particular note is the semiresonant system used in the MADMAX experiment \cite{Caldwell:2016dcw,MadmaxWhitepaper}, where dark matter axions couple to the traveling waves between a series of dielectric plates inside a dipole magnet.  Results from a recent small-scale prototype \cite{garcia2024searchaxiondarkmatter} have demonstrated sensitivities to axion couplings below existing limits from solar experiments.  
    
\subsection{Electromagnetic Searches Beyond Microwave Cavities}
\label{subsec:variants}

\subsubsection{LC Circuit Haloscopes}

While the axion haloscope was originally proposed as a resonator with the same length scale as the axion Compton wavelength, there is no fundamental issue with resonant detection of axion wavelengths exceeding the scale of the detector. One approach which is well-suited to lower frequencies than the microwave cavity searches is to exploit the oscillating effective current induced by the axion
field within a static external magnetic field. The
 ``wire in cavity'' LC circuit concept was proposed in ref.~\cite{PhysRevLett.112.131301} (see also~\cite{Chaudhuri:2014dla} for a related concept for dark photon detection), wherein an oscillating induced effective current in turn
induces a real oscillating magnetic field, which is detected via an induced EMF in a coil.  The resonant frequency can be tuned to the axion mass through an LC circuit or similar method.  Recently a series of experiments have demonstrated the feasibility of the LC circuit technique~\cite{PhysRevD.99.052012,PhysRevLett.127.081801,PhysRevLett.124.241101,Gramolin:2020ict} (ABRACADBRA, ADMX-SLIC, and others)  and a set of experiments with significant reach into unexplored axion parameter space are in preparation. 

{$\bullet$ ABRACADABRA}

\begin{figure}[ht]
\centering
\includegraphics[trim={5cm 17cm 5cm 8cm}, clip, width=0.75\textwidth]{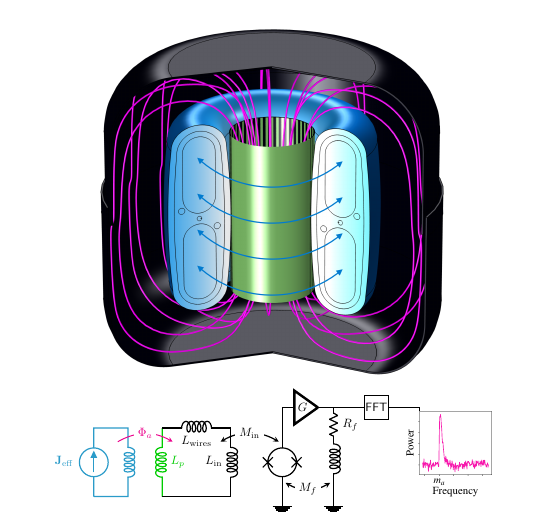}
\caption{
Top: Sketch of an ABRACADABRA prototype showing the
effective ax\-i\-on-in\-duced current (blue), sourced by the
toroidal magnetic field,
generating a magnetic flux (magenta) through the pickup cylinder (green) in the toroid bore.
Bottom: Schematic of the signal readout. The pickup cylinder $L_p$ is inductively coupled to
the axion effective current. The induced current is read out through a dc SQUID
coupled to the circuit through $L_{\rm in}$.
From \citen{PhysRevLett.127.081801}.
}
\label{fig:abracadabra}
\end{figure}

The first realization of the low-mass axion detection scheme was
A Broadband/Resonant Approach to Cosmic Axion Detection with an Amplifying B-field Ring Apparatus
(ABRACADABRA)~\cite{Kahn_2016}. The ABRACADABRA proposal consisted of a toroidal magnet to source the external magnetic field, resulting in an azimuthal effective current,  with the benefit of the physical  oscillating magnetic field signal being in a region of zero background magnetic field~\cite{Kahn_2016}. The experiment could be run in broadband or resonant mode, with the first prototypes taking broadband data  \cite{PhysRevD.99.052012,PhysRevLett.127.081801}.
The recent ``10 cm'' apparatus consisted of a small toroidal magnet (12 cm diameter, 12 cm tall) with peak
field around 1 T. The central magnet bore contains the inductive loop (fig.~\ref{fig:abracadabra}) which is
read out by a dc SQUID at temperature
around 0.5 K \cite{PhysRevLett.127.081801}.
The latest limits were in the 0.41-8.27 neV mass range
for couplings close to and somewhat exceeding the CAST bound~\cite{PhysRevLett.127.081801}.
Future improvements could probe the QCD axion in this mass range.

{$\bullet$ SHAFT}

Another search based on the axion induced current is the
Search for Ax\-i\-on-Like Dark-Mat\-ter with Ferromagnets
(SHAFT)\cite{Gramolin:2020ict}.
Here, a pair of independent toroidal coils are wrapped around high-per\-me\-a\-bi\-li\-ty cores.
A 1.5 T azimuthal magnetic field is applied, and each coil has its own readout; the induced signal is read out by
a dc SQUID. The two-chan\-nel scheme allows for enhanced rejection of electromagnetic backgrounds.
Limits were placed on 12 peV to 12 neV axion masses, with the lightest range below 0.1 neV just exceeding the CAST constraints.
Improvements to the sensitivity by four orders of magnitude could be envisioned,
thereby perhaps achieving sensitivity to QCD axions in this mass range.

{$\bullet$ DM Radio}

To access the lower axion masses at QCD sensitivity, the LC-circuit based DM-Radio~\cite{Silva-Feaver:2016qhh,PhysRevD.106.112003} has been proposed. This is to proceed in several stages.  The first, DM-Radio 50L will demonstrate LC circuit technology on a large scale with a 50 Liter volume and access frequencies as low as 10 kHz, albeit at sensitivity above the QCD axion band.  The DM-Radio m$^3$ is a cubic meter instrumented volume projected to cover the 10 to 200 MHz range \cite{dmradiocollaboration2023electromagnetic} with QCD axion sensitivity.  A future, much larger, DM-Radio-GUT stage with resonant readout is proposed to access frequencies below 10 MHz with sensitivity to the QCD axion.

\subsubsection{Heterodyne Detection}
    In the traditional haloscope detector, the axion field interacts with a static magnetic field to produce a photon at a frequency corresponding to the axion mass $\omega_a$.  If the magnetic field is non-sta\-tic and varying with frequency $\omega$, then photons can be produced at frequencies $\omega \pm \omega_a$.   This suggests the possibility of designing a heterodyne haloscope without the usual large superconducting magnet \cite{GORYACHEV2019,THOMSON2021,Berlin:2019ahk}.  It presents two challenges, however. First, the magnetic fields achievable with oscillating fields are much smaller than those achievable with static magnets, and second, time-vary\-ing magnetic fields tend to also vary spatially, so the overlap of the applied magnetic field, the axion field, and the resultant photon electric field must be carefully aligned.  A proposed technique is to use high-Q superconducting cavities with engineered TM and TE modes such that the field overlap is non-zero for axion conversion when the TM mode is excited with a large-power RF signal \cite{Berlin:2020vrk,Lasenby:2019prg}.  This draws from the experience of high-power RF cavity design for accelerator engineering, and the hope is that the high $Q$ achievable with superconducting cavities will compensate for the smaller applied magnetic field.  The primary benefit is that, because the difference between the TE and TM mode frequency must correspond to the axion frequency, much smaller axion masses can be probed than would normally be accessibly by microwave cavities.

\subsubsection{Broadband Reflectors}
    Ax\-i\-on-to-photon conversion can result from any electromagnetic boundary parallel to an applied magnetic field.  From this point of view, resonant haloscopes can be viewed as conveniently spaced boundaries for a particular axion frequency \cite{Jaeckel:2013eha}.  Alternately, one can cede resonant enhancement and instead focus photons produced from axion conversion at boundaries onto a detector for a broadband search \cite{Horns:2012jf}.  Several experiments have been proposed based on this principle~\cite{FUNKExperiment:2020ofv}, but in general the loss of resonant enhancement makes it difficult to achieve QCD sensitivity.  The BREAD experiment uses a reflector design that is compatible with a high-field solenoidal magnet targeting high mass axions~\cite{BREAD:2021tpx}. This technique has recently achieved sensitivity beyond the CAST constraint~\cite{Hoshino:2025fiz} and may be viable to probe the QCD axion pending the development suitably sensitive single-pho\-ton far infra-red and THz detectors.

\subsection{Axion-Nucleon Coupling Experiments}
\label{subsec:nucleon}
In addition to the axion-photon coupling that has been the focus of this review, several directions have emerged in recent years toward detecting the interactions of axions with nuclei. This includes the `fundamental' interaction of the axion with the CP violating gluon field strength, as well as derivative (gradient) couplings of axions with nuclear spins~\cite{Moody:1984ba, Graham:2013gfa}. While these interactions are particularly difficult to pursue in precision experiments, progress has been made toward several techniques.

One phenomenon, regardless of the makeup of dark matter, is that the ax\-i\-on-nu\-cle\-on coupling would manifest as a macroscopic, spin-de\-pen\-dent ``5th force'', in addition to a Yukawa force in the presence of CP violation~\cite{Moody:1984ba}.  The magnitude of this force varies between axion models, but current torsion pendula experiments are still many orders of magnitude away from the sensitivity necessary to detect even optimistic forces compatible with the QCD axion~\cite{PhysRevLett.106.041801,PhysRevLett.82.2439,PhysRevD.87.011105,PhysRevLett.120.161801}. A dark matter axion field would induce oscillating nuclear electric dipole moments, as well as having a gradient coupling with nuclear spin~\cite{Graham:2013gfa}. Another manifestation is the proposed `piezoaxionic effect' in which QCD axion dark matter generates oscillatory nuclear Schiff moments, which in a piezoelectric crystal can result in strain, or alternatively which can source axion forces~\cite{Arvanitaki:2021wjk,Arvanitaki:2024dev}.
 
\subsubsection{ARIADNE}
The Axion Resonant InterAction Detection Experiment (ARIADNE) is a proposal to detect the effect of the weak short-range force mediated by an axion \cite{PhysRevLett.113.161801}.   The goal is to construct a system with a resonant coupling between a rotating tungsten mass with periodic structure, which sources a slowly oscillating axion field through a Yukawa term (proportional to a small background level of CP violation) and  the NMR frequency of polarized $^3$He nuclear spins. The axion field sources a fictitious magnetic field which is sensed at a very low level in a magnetically shielded system. This strategy has the potential to probe axion masses between $0.1$–$10$ meV  and the technology  for this experiment is currently under development~\cite{ARIADNE:2017tdd,Fosbinder-Elkins:2017osp,Lohmeyeral2020,Gkika:2024zza}.

\subsubsection{CASPEr}
 The Cosmic Axion Spin Precession Experiment (CASPEr) searches for oscillating nuclear spins in the presence of axion dark matter~\cite{Budker2013, Garcon2017,Kimball2017}. There are two directions,  CASPEr-electric and CASPEr-wind. CASPEr-e searches for nuclear spin precession in the background of a large effective electric field in a solid state system. Here, an oscillating  nuclear CP-odd moment (e.g. Schiff moment, nuclear electric dipole moment) arises from the axion-gluon coupling in the background of an oscillating background $\theta$, predicted to be  $\sim 4\times10^{-19}$ if the QCD axion makes up the local dark matter density. This effect in principle has very little model dependence for the QCD axion, although the precise theoretical calculations of the size of the effect in nuclei and atoms have some uncertainties. Recent experiments have demonstrated the technique in the 162-166 neV mass range~\cite{PhysRevLett.126.141802}, currently still well within neutron star and supernova constraints (or alternatively testing a background  $\theta~10^{-6}$). In another approach, no evidence of an oscillating EDM at very low masses (less than $10^{-17}$\,eV) has been seen by comparing the $^{199}$Hg and ultra cold neutron precession frequencies in neutron EDM experiments \cite{PhysRevX.7.041034}, though the sensitivity is still within astrophysical bounds and several orders of magnitude away from the expected QCD axion coupling.  Several orders of magnitude improvement is expected and further theoretical and experimental development is ongoing~\cite{aybas2021quantum,Sushkov:2023nrn,Winter:2024cyp}.
 
 CASPEr-wind searches for an oscillating fictitious magnetic field in the presence of axion dark matter, which arises from the gradient coupling and is generically present but more model-dependent.  A dedicated series of experiments for gradient couplings using nuclear spin co-magnetometers is underway at very low axion masses, ranging from $10^{-22}$-$10^{-12}$\,eV:  CASPEr-Wind ZULF\cite{Wu:2019exd,PhysRevLett.122.191302,doi:10.1126/sciadv.aax4539},  NASDUCK~\cite{doi:10.1126/sciadv.abl8919}, and others. The most recent results have exceeded astrophysical limits for feV and peV mass axions \cite{Lee:2022vvb,Bloch:2022kjm}.  The sensitivity of these searches is still many orders of magnitude away from the QCD axion predictions, and is undergoing development.

\section{Summary and Outlook}
\markboth{SUMMARY and OUTLOOK}{SUMMARY and OUTLOOK}
\label{sec:summary}

\begin{figure}[ht]
\centering
\includegraphics[trim={2.2cm 0 3.5cm 0}, clip, width=1.0\textwidth]{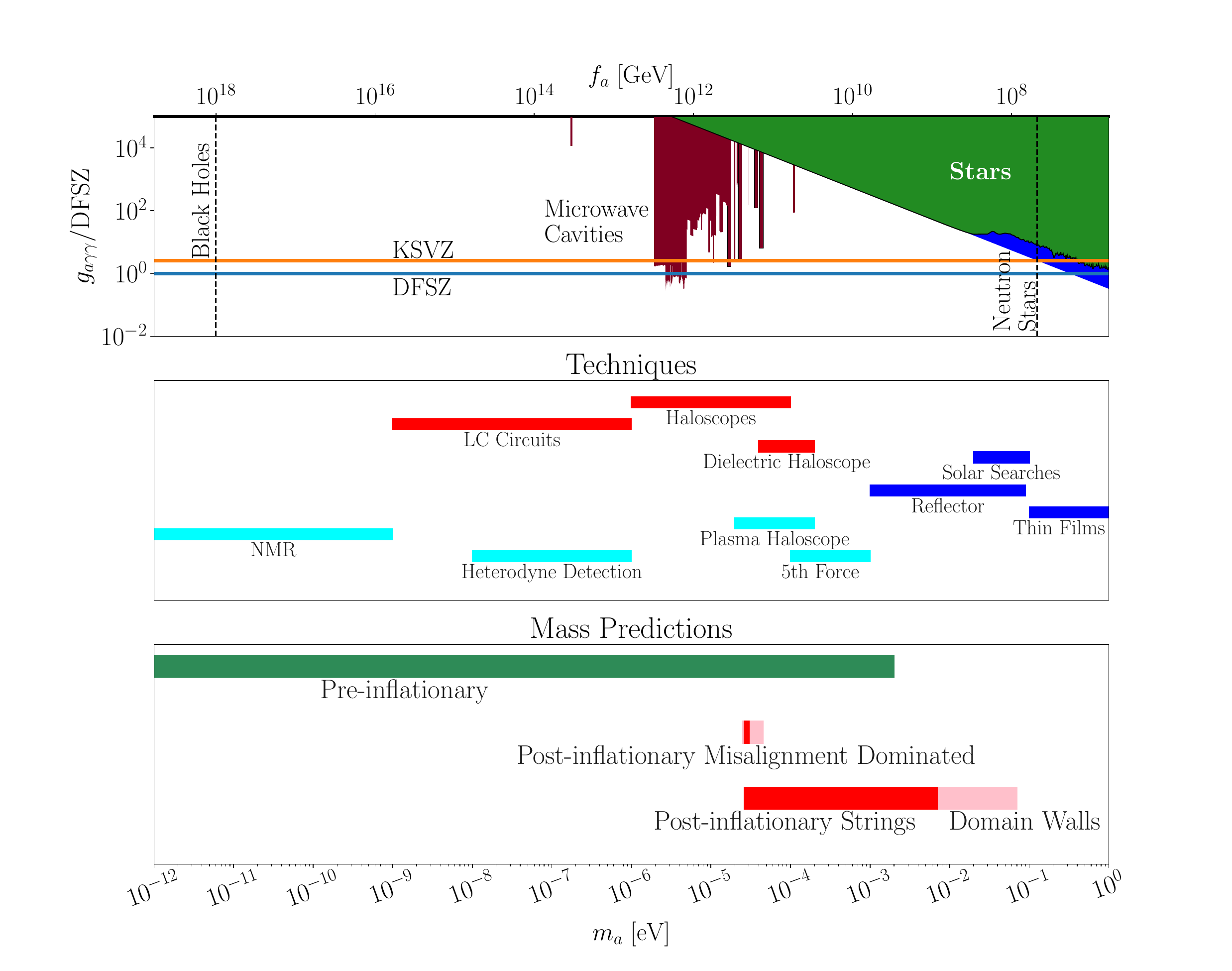}
    \caption{The overall status of searches for the QCD dark-mat\-ter axion.
The horizontal axis is axion mass (lower axis) and axion scale $f_a$ (upper axis).
The vertical axis is the axion to
photon coupling in units of the DFSZ coupling. The upper plot shows the currently achieved
axion mass and photon coupling bounds, with a local dark matter density of 0.45 $\mathrm{GeV}/{\mathrm{cm}^3}$ assumed for haloscopes; see Section~\ref{sec:experiment}. The vertical dashed lines indicate the approximate lower (upper) bounds on the QCD axion from black hole (neutron star) observations; see Section~\ref{sec:astro}. The middle plot indicates where the various
search techniques have optimal sensitivity; see Section~\ref{sec:experiment}. The lower plot shows a high-level
view of mass predictions grouped into pre-in\-fla\-tion\-ary,
post-in\-fla\-tion\-ary (misalignment)
and post-in\-fla\-tion\-ary (ax\-i\-on-string) scenarios; see Section~\ref{sec:cosmo}.}
    \label{fig:triptych}
\end{figure}

Figure \ref{fig:triptych} shows the current status of searches for the QCD dark-mat\-ter axion.
It is clear that we are at the very beginning of the exploration of the QCD axion as a dark matter candidate.
In the near future there are two clear paths that seem most likely to lead to
a discovery.  First, cavity haloscope experiments that have already reached DFSZ sensitivity in the $\mu$eV mass range must be expanded as far and deep as possible, testing a broader mass range as well as our uncertainty in the local dark matter energy density. Second, proposed techniques to explore mass ranges inaccessible to microwave cavity-style experiments must be developed into mature production experiments.

The outlook for improving the mass range of cavity haloscopes has, up until recently, appeared poor: while the axion coupling scales with the mass, the natural cavity size scales as $m_a^{-3}$ and the quantum noise limit scales as $m_a$, suggesting sensitivity should fall as $m_a^{-3}$.  However, several of the technologies we discussed have recently been demonstrated to considerably mitigate this scaling effect.   Multicavity systems can boost the volume scaling by an order of magnitude (or more, limited by mechanical complexity).  Quantum detection techniques: Squeezing, qubit, or
bo\-lo\-met\-ric-based methods could evade the quantum noise limit issue and potentially reduce noise by perhaps an order of magnitude.  Advances in sustaining high-Q for a superconducting cavity in a high magnetic field could increase sensitivity by two orders of magnitude or more, even at higher frequencies.  High-fi\-eld and large-vol\-ume magnets are also being developed, but their realization is unlikely to result in an order of magnitude improvement in practical stored energy for axion experiments in the near future.  The current DFSZ results from ADMX reach up to 4 $\mu$eV with a run plan that extends to 10 $\mu$eV, so we shall take 10 $\mu$eV as a basis for what can be reachable with existing technology (assuming a local dark matter density of 0.45 $\mathrm{GeV}/{\mathrm{cm}^3}$).  Considering sensitivity scaling to degrade as $m_a^{-2}$ with increasing ax\-i\-on mass (including the use of beyond-stan\-dard-quan\-tum-limit detectors previously mentioned), and the (admittedly optimistic) four orders of magnitude from improvements from technologies above, we can speculate that haloscopes could reach close to the meV mass scale at DFSZ sensitivity, completely covering post-in\-fla\-tion\-ary misalignment dominated dark matter ax\-i\-on creation scenarios and the upper part of pre-in\-fla\-tion\-ary axion creation scenarios.  To achieve this will require a unified effort to combine all demonstrated technological innovations into a suite of cavity experiments that span the frequency range.

The remaining axion parameter space, above an meV and below a $\mu$eV, remains to be covered by sensitivity improvements from experiments being currently being developed beyond the cavity haloscopes, or new, as yet-un\-demon\-stra\-ted technologies.  Here the path forward is somewhat less clear. On the high-mass end, the produced photon wavelength is much smaller than natural experiment scales, so it may be that reflector or thin-film style detectors could be scaled, along with far-IR and THz photodetection technology, to be sensitive to DFSZ-coupled axions. A large dedicated solar experiment or improved astrophysical constraints may probe this region as well. Another direction are short-range axion-mediated forces. On the low-mass end, it is quite likely that LC-circuit style haloscopes can be scaled to reach the DFSZ milestone for many orders of magnitude below the current $\mu$eV bounds.  At neV-scale axion masses, the
most-pro\-mis\-ing experimental technologies rely on nuclear couplings via the NMR-type measurements, but early demonstrations are as yet far from QCD axion sensitivity.  It is clear that covering the entire possible QCD axion mass range will require the support of a broad variety of small-scale demonstrators to illuminate which paths are the shortest to a discovery.

In the event that an ax\-i\-on-like sig\-nal is detected, there would immediately be a wealth of information to extract.  In most detector configurations, the kinematic distribution of the local dark matter is derivable from the axion line shape (that is,
the local axion phase space) as well as its evolution over time.  The axion phase space
encodes the infall history of the axion dark matter, which opens the possibility that
the axion discovery yields detailed information on the history of our galactic structure formation. Pre- and post-dis\-covery, the particle physics and phenomenology of the axion will require further study, but would already unambiguously establish a new particle physics scale far above the electroweak scale.  For an axion detection in a haloscope, for example, it is only the product of the ax\-i\-on-pho\-ton coupling-squared and the local dark matter density that is measured; there is no guarantee that this signal comes from a QCD axion instead of, say, an ax\-i\-on-like particle, and there is no way of knowing whether the ax\-i\-on comprises all of the local dark matter or merely a fraction. Further experiments exploring several couplings to fermions, and especially those testing the QCD coupling, will be needed to establish the new particle as the QCD axion and its abundance as the dark matter abundance.

Thus there are considerable theoretical and experimental challenges in the search for the QCD
dark-mat\-ter ax\-i\-on. Nonetheless, there is room for optimism: The ax\-i\-on continues to be a very attractive particle dark-mat\-ter candidate. Sensitivity to the very compelling DFSZ
ax\-i\-on coupling has been demonstrated, and production experiments now operate at that
sensitivity over very promising axion masses. There is renewed appreciation for
QCD ax\-i\-on masses well below and above the microwave cavity QCD ax\-i\-on mass scale,
and technologies are being developed that
may have the requisite sensitivity in these lower and higher masses.
The number of experimenters and experiments are growing, and there continues
to be better understanding of ax\-i\-on phenomenology.
Overall, the case for
the QCD dark-mat\-ter ax\-i\-on is strong and the experimental ax\-i\-on search program
is accelerating. It may be that a discovery is near.

\subsection*{Acknowledgments}
We thank our many colleagues and collaborators for their insights and conversations on QCD axions, and especially David Cyncynates, Ed Hardy, Junwu Huang, Sanjay Reddy, Ben Safdi, Andrew Sonnenschein, Ken Van Tilburg, and Giovanni Villadoro whose exchange of ideas was particularly helpful in the preparation of this review. 
We thank David Cyncynates, Sanjay Reddy, and Giovanni Villadoro for helpful comments on the manuscript and Ciaran O'Hare for maintaining an online repository of recent axion bounds~\cite{ciaran_o_hare_2020_3932430}. M.B. is supported by the U.S. Department of Energy Office of Science under Award Number DE-SC0024375 and the Department of Physics and College of Arts and Science at the University of Washington. G.R. is supported by U.S. Department of Energy Office of Science under Award Number DE-SC0011665.  M.B. is also grateful for the hospitality of KITP and Perimeter Institute, where part of this work was carried out. This research was supported in part by grant NSF PHY-2309135 to the Kavli Institute for Theoretical Physics (KITP). Research at Perimeter Institute is supported in part by the Government of Canada through the Department of Innovation, Science and Economic Development and by the Province of Ontario through the Ministry of Colleges and Universities. This work was also supported by a grant from the Simons Foundation (1034867, Dittrich). 

\newpage

\newpage
\markboth{BIBLIOGRAPHY}{BIBLIOGRAPHY}

\bibliography{main}
\bibliographystyle{JHEP}

\end{document}